\pdfoutput=1
\documentclass[twocolumn]{aastex7}

\usepackage{amsmath}
\usepackage{hyperref}

\usepackage{colortbl}

\newcommand{\nt}{\textcolor{black}}

\begin{document}

\title{The Rocky Planet Picture Show: Implementation of Surface Reflection and Emission in \texttt{POSEIDON} with Application to and Interpretation of JWST Data}

\author[0000-0003-0814-7923]{Elijah Mullens}
\email{eem85@cornell.edu}
\affiliation{Department of Astronomy and Carl Sagan Institute, Cornell University, 122 Sciences Drive, Ithaca, NY 14853, USA}

\author[0000-0003-4816-3469]{Ryan J. MacDonald}
\altaffiliation{NHFP Sagan Fellow}
\affiliation{School of Physics and Astronomy, University of St Andrews, North Haugh, St Andrews, KY16 9SS, UK}
\affiliation{Department of Astronomy, University of Michigan, 1085 S. University Ave., Ann Arbor, MI 48109, USA}
\email{ryan.macdonald@st-andrews.ac.uk}

\author[0000-0002-8871-773X]{Marina E. Gemma}
\email{mgemma@amnh.org}
\affiliation{Department of Astrophysics, American Museum of Natural History, New York, NY 10024, USA}
\affiliation{Department of Earth and Planetary Sciences, American Museum of Natural History, New York, NY 10024, USA}

\author[0000-0001-6092-7674]{Ishan Mishra}
\email{ishanm@ipac.caltech.edu}
\affiliation{Caltech/IPAC, 1200 E California Blvd., MC 100-22, Pasadena, CA 91125, USA}

\author[0000-0002-6385-7672]{Esteban Gazel}
\email{egazel@cornell.edu}
\affiliation{Department of Earth and Atmospheric Sciences, Cornell University, 112 Hollister Drive, Ithaca, NY 14853, USA}

\author[0000-0002-8507-1304]{Nikole K. Lewis}
\email{nkl35@cornell.edu}
\affiliation{Department of Astronomy and Carl Sagan Institute, Cornell University, 122 Sciences Drive, Ithaca, NY 14853, USA}
 
\begin{abstract}
\noindent The surface characterization of rocky exoplanets via emission spectroscopy represents a frontier of current (JWST) and future (HWO) observational efforts. Here, we implement new features in the open-source retrieval code \texttt{POSEIDON (v1.4)} to fully account for an emitting and reflecting planetary surface and an overlying absorbing and scattering atmosphere. We show that realistic rocky surfaces (with wavelength-dependent albedos derived from laboratory measurements) affect emission spectra by imparting mid-infrared diagnostic absorption features, imprinting pseudo-features due to atmospheric transparency windows, and flipping absorption features to emission via surface-atmosphere interface \nt{pseudo-}temperature inversions. We demonstrate that current JWST spectral data can distinguish between tenuous (low surface pressure, \nt{$\leq$ 1 bar}) and thick (high surface pressures, \nt{$\geq$ 0.1 bar}) atmospheres by performing atmosphere + surface retrievals on published JWST emission data of the rocky worlds TOI-1685b and 55~Cancri~e. We then explore JWST MIRI LRS's capability to constrain surface geology of rocky worlds, finding that with sufficient SNR retrievals can distinguish between granite-like and basaltic surfaces \nt{for synthetic datasets}. Finally, we provide an open-source database of lab-derived surface albedos (in the form of directional-hemispherical reflectances), organized by geologic classification and include supplemental tables developed to foster future collaboration between geology and exoplanet science. Our atmosphere + surface retrieval technique provides a pathway to probe geologic processes on rocky exoplanets, showing that upcoming JWST data for terrestrial worlds will enable a deeper exploration of rocky surfaces beyond our Solar System. 
\end{abstract}


\section{Introduction} \label{sec:intro}

% Due to their thin and tenous atmospheres when compared to large-scale height gas-giants, the exploration of rocky worlds has been mostly limited to ones orbiting dwarf stars by leveraging the rocky world's close proximity to their host star and large signal-to-noise ratios in transmission and emission measurements. However, it remains a mystery of whether or not rocky worlds around dwarf stars can retain an atmosphere due to the strong UV emission and flares of their host stars.

The characterization of terrestrial exoplanets ($R_p \leq 1.8 R_{\Earth}$, with a well-defined solid surface) using spectral information represents the frontier of current and future observational capabilities. The atmospheres of rocky worlds are expected to be extraordinarily diverse with differing surface pressures, gas-phase constituents, and aerosols \citep[see][for a review]{Wordsworth2022}. Atmospheric diversity is evident even within our own Solar System: Mercury has a tenuous exosphere with a surface pressure $< 10^{-14}$\,bar \citep{Domingue2007,Wurz2025}; Venus has a CO$_\mathrm{2}$-dominated atmosphere with high-altitude thick sulfuric acid (H$_2$SO$_4$) clouds and a surface pressure of 100 bars \citep{Taylor2018,Dai2025}; Earth has an N$_2$-O$_2$ atmosphere with patchy H$_2$O clouds and a 1 bar surface pressure \citep{Pollack1980}; and Mars has a tenuous CO$_2$-dominated atmosphere with seasonal CO$_2$ and H$_2$O clouds and a 10 mbar surface pressure \citep{Moroz1976,Nakagawa2019}. The atmospheric composition and surface pressure of the four Solar System rocky planets give vital clues to the evolutionary processes shaping each world, including atmospheric escape, solar insolation, surface-atmosphere interactions, geologic evolution, and, in the case of Earth, life \citep[see][for a review]{Lammer2018}. Exoplanet atmospheres and surfaces can be characterized via the techniques of transmission, emission, and reflection spectroscopy \citep[see][for a recent review of JWST transmission and emission observations of rocky planets]{Kreidberg2025}.

Transmission spectroscopy measures light attenuated by the atmosphere at the day-night terminator of a planet. For rocky exoplanets,  many JWST transits are required to build a sufficient signal-to-noise ratio to detect atmospheric signals analogous to the four Solar System terrestrial planets, even for planets orbiting bright dwarf stars. Models predict that 20-100 JWST transits are needed to detect the attenuation of starlight from rocky exoplanet atmospheres \citep{MacDonald2019,Kaltenegger2020,Lin2021,Gomez2023}, with recent JWST observations inferring potential hints of atmospheres on rocky worlds with $<$ 5 transit observations for LHS~1140b \citep{Damiano2024,Cadieux2024} and TRAPPIST-1e \citep{Espinoza2025,Glidden2025}. Non-detections of atmospheres via transmission spectra, however, are degenerate with multiple interpretations: i) a bare rocky surface, ii) optically-thick high-altitude clouds, iii) a high atmospheric mean molecular weight, and/or iv) a thin atmosphere with insufficient SNR to detect molecular absorption \citep{Mayorga2021,LustigYaeger2023}. Additionally, stellar contamination due to unocculted active regions can imprint features in transmission spectra that mimic an atmosphere \citep[e.g.,][]{Rackham2018,Moran2023}. Since transmission spectroscopy measures atmospheric attenuation of starlight, it cannot directly detect the composition of a planet's surface.

Secondary eclipse observations directly measure the planetary flux (reflected starlight + thermal emission) of an exoplanet's dayside as it eclipses behind its host star. The presence or absence of rocky planet atmospheres can be probed via infrared secondary eclipse observations, be they photometric \citep[e.g.,][]{Mansfield2019,Koll2019,Durcot2025} or spectroscopic  \citep[e.g.,][]{Morley2017,Malik2019,Hu2024}. Photometric thermal emission can directly measure the dayside temperature of rocky worlds. By comparing the expected equilibrium dayside temperature and the observed temperature, one can infer whether or not a sufficiently thick atmosphere exists to redistribute heat from the dayside to the nightside hemisphere \citep[e.g.,][]{Green2023}. Spectroscopic thermal emission observations additionally can contain atmospheric signatures, such as gas-phase or aerosol absorption and emission features, which both confirm the existence of an atmosphere and probe the pressure-temperature profile \citep[e.g.,][]{Malik2019}. Crucially, emission spectra can also contain absorption features originating from the surface that depend on the intrinsic geology and surface composition \citep{Hu2012,Fortin2022,Paragas2025,First2025}.

%Without a thick atmosphere, it is expected that a planet's surface emission will dominate the observed signal. Surfaces themselves can impart diagnostic absorption features in spectra that are indicative of specific geologies \citep[e.g.,][]{Hu2012,Paragas2025,Hammond2025}.

Reflected light also offers the promise of detecting atmospheres and surface properties on rocky exoplanets. Detecting reflected light from rocky worlds is at the forefront of next-generation telescope development (e.g., the Habitable Worlds Observatory, \citealt{Decadel2021,East2024}), where these observations can prove to be a powerful probe of atmospheric composition and surface composition \citep[][]{Feng2018,Carrion-Gonzalez2020,Damiano2020,GomezBarrientos2023,GoodisGordin2024,Cowan2025,Wakeford2025,Hu2025,KrissansenTotten2025, Zelakiewicz2026}. Reflected light can also be leveraged to detect the existence of tenuous (Mars-like) atmospheres with weak, difficult-to-detect gas-phase absorption by measuring a higher than expected reflected-light albedo from high-altitude, high-albedo clouds \citep{Mansfield2019}. While some JWST observations are beginning to probe wavelengths where reflected light can contribute to the observed infrared spectra \citep[e.g.,][]{Luque2024}, detecting atmospheric features via reflected light of these smaller objects continues to prove difficult.

%\citet{Mansfield2019} argue that more tenuous atmospheres that contain high-altitude, high-albedo clouds (akin to Mars) could be inferred by measuring a high albedo (via reflected light), as well as directly detecting aerosol absorption and emission features. 

%Pre-JWST, exoplanet emission measurements were primarily made using the Spitzer/IRAC channel 2 (4.5 \textmu m) photometric channel \citep[e.g.,][]{Kreidberg2019,Crossfield2022}. 
JWST's mid-infrared instruments are now detecting dayside thermal emission from a multitude of rocky exoplanets. The photometric MIRI F1500W filter (15 \textmu m) has been used to determine the dayside brightness temperature of TRAPPIST-1b \citep{Green2023, Ih2023}, TRAPPIST-1c \citep{Zieba2023}, \nt{LHS 1478b \citep{August2025}, TOI 1468b \citep{MeierValdes2025}, LHS 1140c \citep{Fortune2025},  LTT 3780b \citep{Allen2025}, GJ 3929b \citep{Xue2025}, and GJ 3473 b \citep{Holmberg2026}}. The spectroscopic MIRI/LRS instrument (5-12 \textmu m) has measured dayside emission spectra for GJ 367b \citep{Zhang2024}, GJ 486b \citep{Mansfield2024}, 55 Cancri e (55 Cnc e) \citep{Hu2024}, GJ 1132b \citep{Xue2024}, and LTT 1445Ab \citep{Wachiraphan2024}. Near infrared instruments have also been used, such as JWST/NIRCam photometry at 2.1 and 4.5 \textmu m and F444W grism spectroscopy (3.94-4.49 \textmu m) of 55 Cnc e \citep{Patel2024, Hu2024}, and JWST/NIRSpec G395H spectroscopy (3.823-5.172 \textmu m) of TOI-1685b \citep{Luque2024} and TOI-561b \citep{Teske2025}. With the exceptions of 55 Cnc e --- for which some visits favor a thick CO/CO$\mathrm{_2}$ atmosphere (\citealt{Hu2024}, but see also \citealt{Patel2024}) --- TOI-561b --- which supports a thick volatile-rich atmosphere \citep{Teske2025} --- \nt{and HD 3167b -- which has evidence of a thick atmosphere cooling the dayside \citep{Coy2026}}, most thermal emission data has been consistent with a blackbody or dark albedo surface, ruling out thick atmospheres. 

\begin{figure*}[ht!]
    % \vspace{-1.3cm}
    \includegraphics[width=1.0\textwidth]{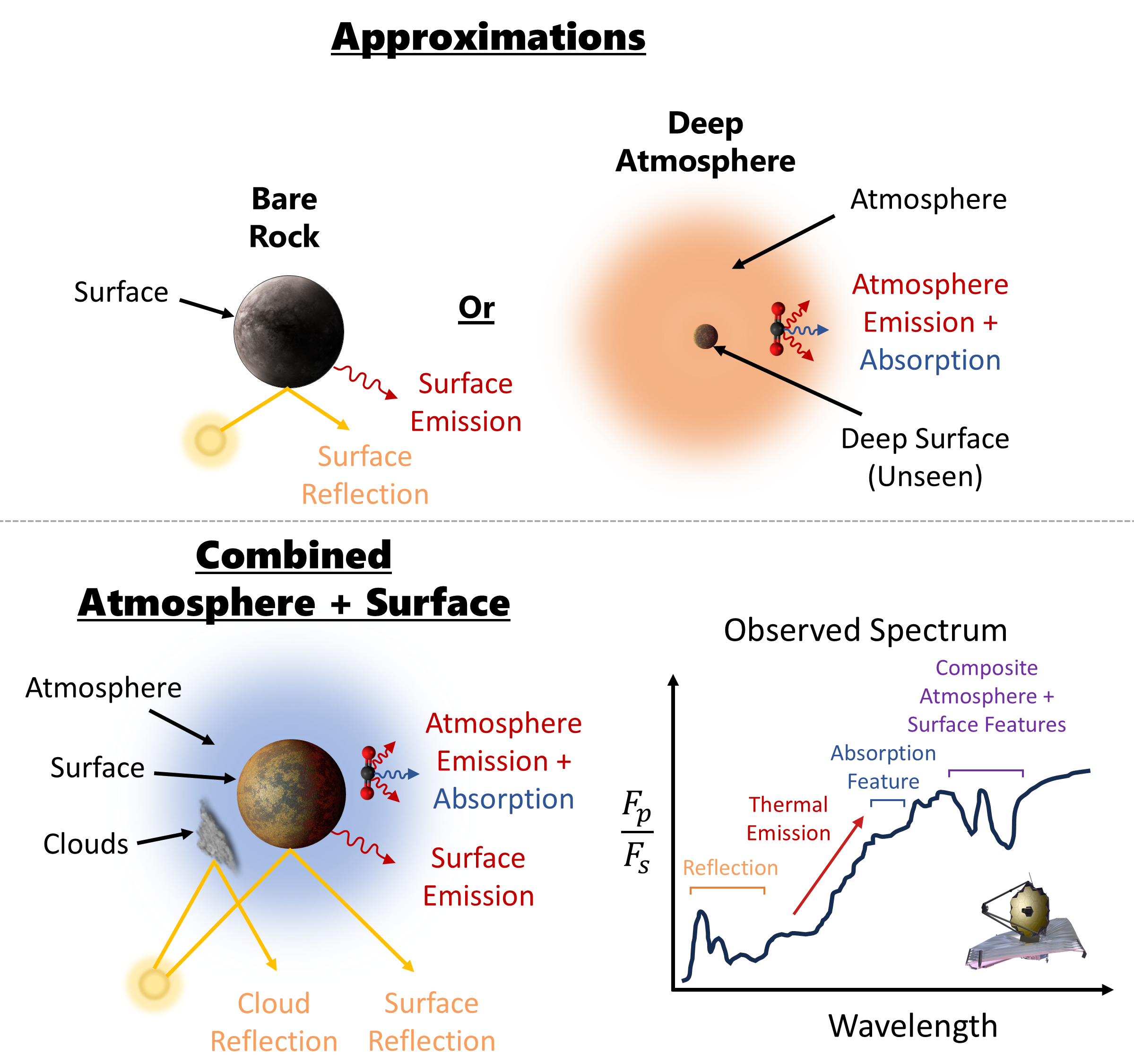}
    \caption{Rocky exoplanet emission spectra has often been modeled and fit with bare-rock and thick-atmosphere approximations, but not simultaneously. \texttt{POSEIDON} has been updated to self-consistently compute the radiative transfer for objects with a thermally emitting and reflecting surface (with multiple components), an overlaying gas-phase atmosphere, and patchy Mie scattering clouds. The absorptive, emitting, reflective, and scattering properties of all three combine to produce the observed spectrum. This allows for a retrieval analysis to let data guide whether a bare-rock, thick-atmosphere, or an intermediate model best describes the data.}  
    \label{fig:first-figure}
\end{figure*}

Previous modeling efforts to fit JWST thermal emission data have heavily relied on simple models that assume either a bare rocky surface or a thick atmosphere. Such forward models have been used to determine whether or not the observed temperature and spectra of the dayside indicates the presence or absence of an atmosphere. For photometric observations, this involves an analysis of the observed dayside brightness temperature that includes using the heat redistribution `$f$-factor' (see \citealt{Wachiraphan2024} for a recent review). For spectral analyses, this involves fitting observed spectra with either a bare-rock model or a thick-atmosphere model and performing a chi-squared analysis or retrieval \citep[e.g.,][]{Luque2024}. 

A more realistic model of rocky exoplanet emergent spectra involves the full interaction of light with the surface and the overlaying atmosphere and clouds (Figure \ref{fig:first-figure}). It is well-known from planetary science that the emergent flux from the Solar System's terrestrial planets is a strong function of surface composition, atmospheric density, and the atmospheric and aerosol composition. In particular, Mercury's flux is dominated by surface geology, Venus's flux is dominated by atmospheric absorption and global reflective clouds, while Earth's flux is controlled by the surface, gas-phase, and clouds \citep[e.g.,][]{Head2007,Selsis2008}. \nt{
Several advanced modeling frameworks exist to self-consistently model rocky planet spectra, including \texttt{HELIOS} \citep{Malik2017,Malik2019,Malik2019b,Whittaker2022}, the Planetary Spectrum Generator \citep{Villanueva2018}, and work done by the Virtual Planetary Laboratory \citep[e.g.,][]{Gu2021}. Codes like this can iteratively solve for the pressure-temperature profile, given atmospheric and surface properties, and can compute spectra from disc-resolved forward model planets. These tools are useful for forward model analyses probing the direct connection between underlying physical assumptions and emergent spectra, but lack the capability to sweep large parameter spaces of atmospheric and surface properties.}

In this work, we develop an inverse (`retrieval') analysis model for rocky exoplanets that couples the thermal emission and reflected starlight contributions from a surface, an atmosphere, and aerosols. In \S \ref{sec:2}, we extend the open-source \texttt{POSEIDON}\footnote{\url{https://github.com/MartianColonist/POSEIDON}} \citep{MacDonaldMadhusudhan2017, MacDonald2023, Mullens2024, Wang2025, Mullens2025} code to include a reflecting and emitting surface in tandem with an atmosphere and patchy clouds (Figure \ref{fig:first-figure}). In \S \ref{sec:3}, we demonstrate that including a reflecting and emitting surface changes many common assumptions about emission spectra compared to `gray' opaque surface. In \S \ref{sec:4}, we apply our new  retrieval model to the JWST emission spectra of TOI-1685b and 55 Cnc e. In \S \ref{sec:5}, we show that stacked JWST MIRI eclipses for rocky planets with no or thin atmospheres can distinguish between different surface materials, which opens the door for surface exogeology of these rocky worlds. Finally, in \S \ref{sec:last}, we summarize our findings, discuss the implications, and provide links to our curated database of surface albedo data and open-source tutorial notebooks. Additionally, in the Appendix we include relevant derivations for bare rocky surface spectra in \S\ref{Appendix-1}, and supplementary geology tables alongside our full surface albedo database in \S\ref{Appendix-2}. 

\section{Enhancements to \texttt{POSEIDON}}\label{sec:2}

Here, we describe the improvements and new features implemented into the open-source atmospheric retrieval code \texttt{POSEIDON} for \texttt{v1.4}. We note that a beta version of the code introduced here was utilized to generate forward model observed reflection and emission spectra of an exo-Europa around a white dwarf in \citet{Mullens2025-2}.

\subsection{Surface Reflection and Emission Implementation}\label{sec:2.1}

The emission and reflection modules in \texttt{POSEIDON} (both the single-stream emission prescription described in \citet{Coulombe2023} and the thermal-scattering and reflection \nt{radiative transfer} prescription adapted from \texttt{PICASO} \citep{Toon1989,Batalha2019,Mukherjee2023} and described in \citealt{Mullens2024}) have been updated to include a surface-layer with a wavelength-dependent albedo (with a surface emissivity ($\epsilon_{\mathrm{surface}}$) of 1.0 - surface albedo)\footnote{Reflective surfaces were first introduced in \texttt{PICASO 1.0} \citep{Batalha2019} (see `Surface Reflectivity' tutorial in \href{https://natashabatalha.github.io/picaso/tutorials.html}{\texttt{PICASO} tutorials}) and fully integrated with \nt{atmospheric radiative transfer in} \texttt{PICASO 4.0} \citep{AdamsRedai2025,Mang2026}. \nt{For more details on the radiative transfer implementation in \texttt{POSEIDON}, see \texttt{emission.py} available on \href{https://github.com/MartianColonist/POSEIDON}{\texttt{POSEIDON}'s github}}}. In emission, the thermal emission of the surface layer is the blackbody spectrum given by the pressure-temperature profile at the surface layer multiplied by the surface emissivity. The surface layer is described by a surface pressure parameter (log$_{\mathrm{10}}$ P$_{\mathrm{surf}}$), where any pressures higher than the surface pressure are opaque and pressures lower are the atmosphere. These models work with clear atmospheres composed of soley gas-phase species, as well as cloudy atmospheres with patchy (or homogeneous) Mie-scattering aerosols. 

We have included three different surface models: a blackbody `gray' surface (which has an constant albedo of 0), a constant albedo surface (which is parameterized by a constant with wavelength albedo, albedo$_\mathrm{surf}$), and a laboratory data surface which utilizes lab-measured wavelength vs albedo spectra (for lab data pre-included in this release of \texttt{POSEIDON}, see \S \ref{sec:database} and \S\ref{Appendix-2}). The lab data surface can be initialized with a single or multiple surface components. If multiple surface components are defined, each component is assigned a surface-coverage percentage (which can either be defined in linear space from 0--1 or in log space) that are pre-normalized to ensure the component percentages add to 1. There are two methods by which to apply the surface component percentages. The first applies the weights to the albedos themselves and computes a single spectrum assuming the weighted albedo (a single instance of radiative transfer is performed). The second computes individual spectra for each surface assuming each surface component is 100\% the surface, and then applies the weights when summing the resultant spectra (N spectra generated for N surfaces). In this work, we opt to use the second method for weighing surface components \nt{in \S \ref{sec:5} due to this method better mirroring the philosophy of previously utilized patchy cloud 1+1D models \citep{Marley2010, Morley2014, Mullens2024},} but note that both are included in \texttt{POSEIDON}, \nt{and that the method of applying weights before computing radiative transfer was utilized in the \texttt{POSEIDON} reflected light retrievals presented in \citet{Zelakiewicz2026} for computational efficiency. The two methods are shown to be equivalent in our online documentation\footnote{\label{fn:surfaces}For a tutorial on how to implement surface models (both with and without atmospheres) in \texttt{POSEIDON}, see the \nt{`Rocky Planets with Reflecting and Emitting Surfaces'} tutorial in \href{https://poseidon-retrievals.readthedocs.io/en/latest/content/forward_model_tutorials.html}{Forward Model Tutorials}.}.}

\subsubsection{Bare Rocky Surface Spectra}

We have also included a model that is composed of solely an emitting and reflecting surface, with no overlaying atmosphere. All the aforementioned surface models (blackbody, constant albedo, and lab data) in the previous section work with this model. These models have a free surface temperature parameter ($T_{\mathrm{surf}}$). We note that in this work we are not modeling 2D surface emission where energy balance is utilized to solve latitude-longitude specific surface temperatures (wherein the sub-stellar point is the hottest and brightest point, \citealt{Hu2012,Hammond2025}), and instead use a free surface temperature parameter removed from energy balance (where we assume that the full observed disc of the planet is a singular surface temperature). \citet{Paragas2025} show that models with a homogeneous surface temperature can sufficiently be compared to more complex models that independently solve for latitude-longitude dependent surface temperatures; \nt{additionally, \citet{Schwartz2015} show that the single-temperature approximation produces near-identical flux at thermal infrared wavelengths that we are focused on fitting in this work.} We choose to not utilize energy balance in our models in order to 1) let the data best fit the derived temperature and 2) account for potential processes (i.e., tidal heating, solid-state greenhouse and anti-greenhouse effects, \citealt{Driscoll2015,Lyu2025}) that could heat the surface hotter (or colder) than energy balance derives. Future work (see \S \ref{sec:6.1}) will explore the inclusion of energy balance and 2D surface emission modeling, \nt{especially as it can produce observable effects in optical wavelengths at higher temperatures ($>$ 1000K, \citealt{Schwartz2015}), and potentially at thermal wavelengths due to temperature-dependent surface emissivities \citep{Thompson2021,Fortin2024}.}

The emergent thermal emission flux from a bare rocky surface is given by

\begin{equation}
    F_{\mathrm{thermal,rock}} = \pi \; B(T_{\mathrm{surface}},\lambda) \; \epsilon_{\mathrm{surface}}(\lambda)
    \label{eq:1}
\end{equation}

\noindent where $B(T,\lambda)$ is the Planck function. Note that because we are only observing one hemisphere of the planet, $\pi$ is used instead of 2$\pi$ \nt{(i.e., the factor of $\pi$ is from integrating over one hemisphere, see \S \ref{Appendix-1})}. This is converted to observed flux via: 

\begin{equation}
    F_{\mathrm{thermal,rock,obs}} = \left(\frac{R_p}{d}\right)^2 \; F_{\mathrm{thermal,rock}}
\end{equation}

\noindent where $R_p$ is the planet's radius and $d$ is the distance to the system. For reflected flux, the geometric albedo ($A_g(\lambda)$) of a bare rock is computed using the 5 Gauss angle integration scheme \nt{(which assumes spherical symmetry)} for a single layer from \citet{Batalha2019}, which is then converted to observed flux via

\begin{equation}
F_{\mathrm{reflected,rock,obs}}(\lambda) = A_g(\lambda) \; \left(\frac{R_p}{a_p}\right)^2 \;  \left(\frac{R_s}{d}\right)^2 \; F_s(\lambda) \; 
\end{equation}

\noindent where $a_p$ is the orbital distance from the star to planet, $R_s$ is stellar radius, and $F_s$ is the stellar flux. For a more in-depth derivation of the bare-rock equations, refer to \S \ref{Appendix-1}\footref{fn:surfaces}.

\subsection{Other New Features}

In this section, we list other improvements we have added to \texttt{POSEIDON} that are related to surface modeling. 

\subsubsection{Shiny Opaque Deck}

While gas giants are not thought to have hard surfaces, they can have opaque cloud decks that act as the deepest observable layer of the atmosphere (i.e., similar to the observable cloud deck of Jupiter). These cloud decks are typically modeled as gray, infinite opacity absorbers. We have updated all cloud models in \texttt{POSEIDON} that include a gray deck (MacMad17 Deck, as well as all the Mie cloud models with a gray deck described in \citealt{Mullens2024}) to optionally make them `shiny' opaque decks described by a constant albedo (albedo$_{\mathrm{deck}}$). This is similar in concept to the constant albedo surface described in \S \ref{sec:2}\footnote{For a tutorial on how to implement gray shiny decks in \texttt{POSEIDON}, see the \nt{`Advanced: Shiny Gray Decks in Eclipse Geometry'}  tutorial in \href{https://poseidon-retrievals.readthedocs.io/en/latest/content/forward_model_tutorials.html}{Forward Model Tutorials}.}.

\subsubsection{Module Rework for HWO}\label{sec:2.2.2}

Looking forward to the launch of the Habitable Worlds Observatory (HWO), which will be observing primarily reflected light from Earth-like planets with little thermal emission contributing to the observed spectra, we have reworked the modules in \texttt{POSEIDON} to allow for only reflected light to be computed (without any thermal components). This rework speeds up forward model computation considerably in a retrieval framework since reflection, which was initially included in the release of \texttt{POSEIDON V1.2} \citep{Mullens2024}, by default, had to be computed with both thermal scattering and reflection. 

As a proof-of-concept, we generated a synthetic HWO Earth-like spectrum by scattering data with HWO's predicted SNR ($\approx$10 with R = 100, errorbar = 3e-11 ppm in Fp/Fs, originally from the \href{https://seec.gsfc.nasa.gov/News_and_Events/Spectral_Retrieval_Tutorial_2024.html}{HWO Retrieval Workshop}) on a \texttt{POSEIDON}-generated Earth-like proxy spectrum and ran retrievals with only reflected light, finding that \texttt{POSEIDON} will be well-suited to perform retrievals on future HWO datasets. We have included this spectrum and retrieval in the appendix (Figure \ref{fig:HWO})\footnote{For a tutorial on how to implement HWO-like reflection spectra in \texttt{POSEIDON}, \nt{following the modeling done in Figure \ref{fig:HWO} and \citet{Zelakiewicz2026},} see the \nt{`Simulating Habitable Worlds Observatory Reflection Spectra'} tutorial in \href{https://poseidon-retrievals.readthedocs.io/en/latest/content/forward_model_tutorials.html}{Forward Model Tutorials}.}. 

See \citet{Zelakiewicz2026} for a more in-depth application, where they simulate observations of Earth-like planets as observed by the HWO Exploratory Analytical Case 5 (EAC5). They model observations using physically motivated noise sources for the 10-m EAC5 design, including constraints such as the coronagraphic throughput, optical throughput, and detector efficiencies. \texttt{POSEIDON} is then utilized to quantify HWO's ability to constrain surface properties alongside atmospheric properties (i.e., gas-phase chemistry and cloud coverage). 

\subsubsection{CLR Prior for Surfaces}

We have included the option for retrievals with multiple surface components to use the permutation-invariant centered-log-ratio (CLR) prior \citep{Benneke2012} on the surface component percentages. The CLR prior has been used in multigas retrievals of sub-Neptunes and rocky worlds where any of the included gases can be the dominant gas (`bulk gas species') \citep[e.g.,][]{Piaulet2024}. This prior is useful when fitting surface component percentages where any of the surfaces are equally likely to be the dominant surface component, and is also a useful prior for ensuring that the retrieved percentages add up to 1 (i.e. 100\%) by pulling random samples for $n-1$ components and computing the final surface component percentage by subtracting the the sampled percentages from 1. In contrast, the uniform prior pulls random samples and then normalizes them to 1. We feature CLR retrievals of surface component percentages in \S \ref{sec:5}. 

\subsection{Surface Albedo Database}\label{sec:database}

We have curated a collection of surface albedos to be pre-included in \texttt{POSEIDON}. In order to elucidate the minerals and rock types included in the database, \S \ref{Appendix-2} includes supplementary geology tables (Tables \ref{table:Geology Glossary}, \ref{table:Surface Albedos Categories}, \ref{table:Surface Albedos Categories 2}) in order to fully contextualize the Solar System and potential exoplanet context of the lab data in Table \ref{table:Surface Albedos}. This collection includes albedos that cover wavelengths regions where reflection and thermal emission dominant (UV-VIS-NIR-MIR) that can be directly applied to HST and JWST data. When possible, all surface albedos are standardized to the form of lab-measured directional-hemispheric reflectivities ($r_\mathrm{dh}$), which is the most common measurement made by commercial instruments to measure material reflectance \citep{Hapke2012} (all derivations in \S \ref{Appendix-1} are based on surface albedos being directional-hemispherical reflectance). 

Our database includes the spectral library from \citet{Hu2012} that contained measurements of eight powdered samples of representative surfaces that has been used extensively in extant analysis \citep[e.g.,][]{Luque2024} with directional-hemispheric reflectivities sourced from \texttt{PLATON}\footnote{https://github.com/ideasrule/platon} \citep{Zhang2019,Zhang2025}. We have also included the \citet{Paragas2025} textural library that expanded on the original \citet{Hu2012} spectral library, with directional-hemispheric reflectivities also sourced from \texttt{PLATON}. We have included the \citet{Hammond2025} albedos curated from the \texttt{RELAB Spectral Database}\footnote{https://sites.brown.edu/relab/relab-spectral-database/} \citep{Milliken2021}, where directional-hemispherical reflectivites are derived from their reported single-scattering albedos ($\omega$)\footnote{https://zenodo.org/records/14017134} via 

\begin{equation}
r_\mathrm{dh} = \frac{1-\gamma}{1+2\gamma\mu_0} \; , 
\end{equation}

\noindent where $\gamma = \sqrt{1-\omega}$ and we take $\mu_0$ = 1 \nt{due to \texttt{POSEIDON} only being capable of modeling secondary eclipse at the time of this work}. We have also included the natural basalt library of \citet{First2025}\footnote{https://zenodo.org/records/12822668} (Table \ref{table:Basalt-Library}), and the lava world surface library of \citet{Fortin2022}\footnote{https://zenodo.org/records/6323322} (Table \ref{table:Lava-Library}), which can be utilized to further explore the geology of rocky worlds. 

%We note that while this work is focused on the inclusion of surfaces in a self-consistent retrieval framework, the actual geological interpretation and application of different surface albedos is pertinent and should be met with care when interpretating any retrieval results (see \S \ref{sec:5}). 

We have also included surface albedos that primarily cover only short wavelengths (UV-Vis-NIR) where reflection dominants (Table \ref{table:Surface Albedos Reflections}) which can be applied to future HWO retrieval analysis. This includes the database compiled by \citet{GoodisGordin2024} which has Earth-like surfaces (forest, grass, snow, etc) and biota from the \texttt{NASA JPL ECOSTRESS Spectral Library}\footnote{https://speclib.jpl.nasa.gov/} \citep{Baldridge2009,Meerdink2019}, \texttt{USGS
Spectral Library}\footnote{www.usgs.gov/labs/spectroscopy-lab/usgs-spectral-library} \citep{Kokaly2017}, and \citet{Sparks2021}. Unlike the albedos above, the albedos in this database are not necessarily directional-hemispherical reflectance, which we make note of in Table \ref{table:Surface Albedos Reflections}. \nt{Additionally, we have included the surface data utilized to model modern Earth for HWO simulations from \citet{Zelakiewicz2026}, which are further detailed in that work. \nt{We note that user-inputted albedos can be easily added to \texttt{POSEIDON}, where the albedo database introduced in this work represents only the datasets that are preincluded and documented in this work.\footnote{For plots of all lab-measured surface albedos vs. wavelength, see the \nt{`Surface Albedo Database'} page in \texttt{POSEIDON}'s \href{https://poseidon-retrievals.readthedocs.io/en/latest/content/opacity_database.html}{Opacity Database}.}} }

\section{Effects of Emitting \& Reflecting Surfaces on Atmospheric Inferences}\label{sec:3}

\begin{figure*}[ht!]
    % \vspace{-1.3cm}
    \includegraphics[width=1.0\textwidth]{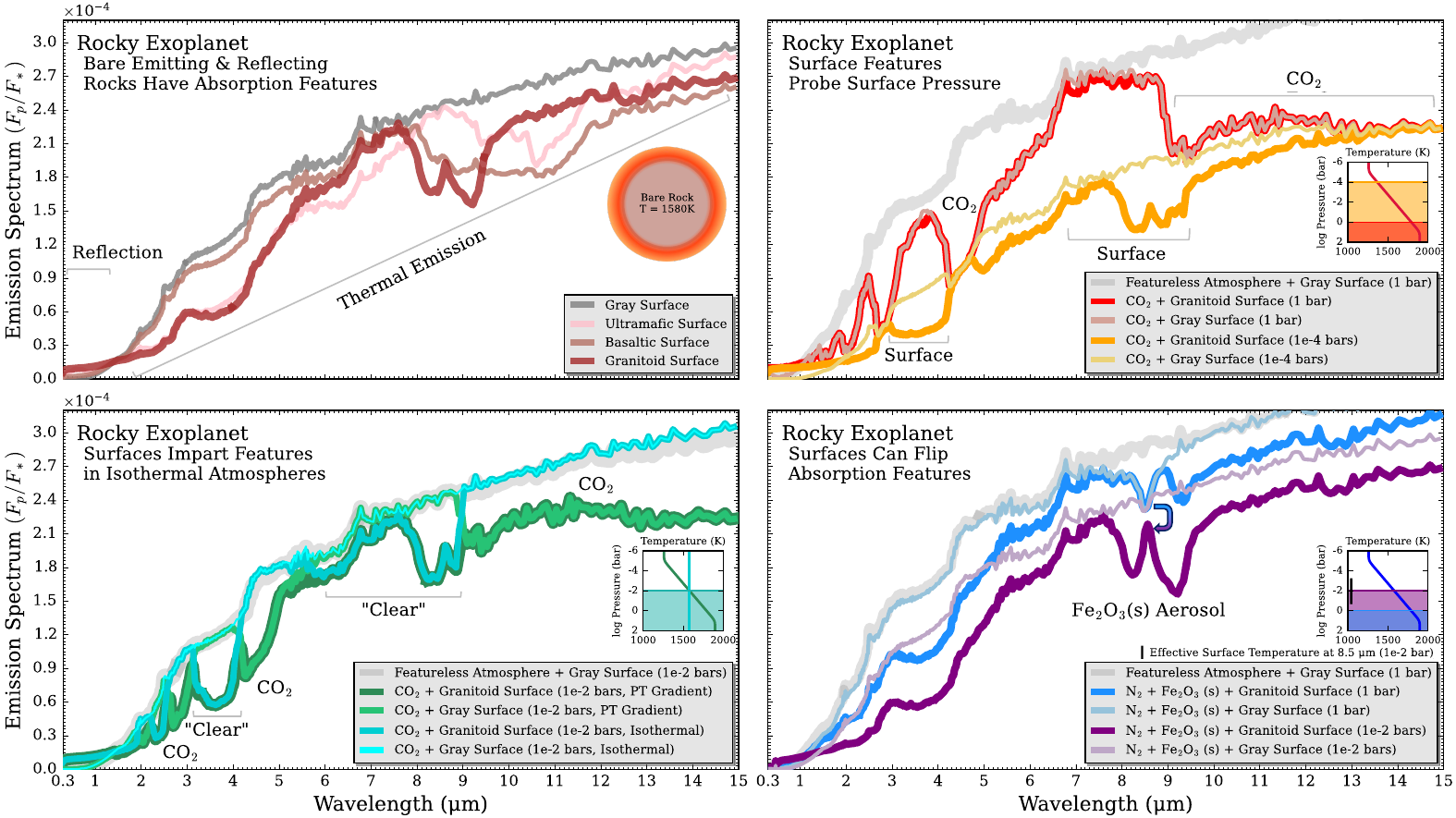}
    \caption{Main takeaways for why including surfaces with wavelength-dependent albedos in radiative transfer models influences resultant secondary-eclipse spectra and atmospheric inferences. \textbf{Top Row} \textit{Left}: Bare rocky surface exoplanets without atmospheres can impart compositionally diagnostic features in spectra due to their wavelength-dependent albedos. \textit{Right}: \nt{Thicker} atmospheres (1 bar) have spectra dominated by absorption from the atmosphere whereas thinner atmospheres (1e-4 bar) have spectra dominated by surface features. This interplay helps constrain surface pressure. \textbf{Bottom Row} \textit{Left}: Surfaces can impart pseudo-features in atmospheres with isothermal pressure-temperature profiles due to atmospheric transparency windows. \textit{Right}: Surfaces can flip atmospheric absorption features into emission features in wavelength regions where atmospheric opacity overlaps high surface albedo, resulting in a surface-atmosphere interface \nt{pseudo-}temperature inversion that effects the spectrum in low surface pressure scenarios \nt{(i.e.,  the intensity from the surface is lower than the source function of the atmosphere at wavelengths where the surface emissivity is low)}. For details on the models, see \S \ref{sec:3}.}    
    \label{fig:main-takeaways}
\end{figure*}

The inclusion of a thermally emitting and reflecting surfaces in radiative transfer is essential when generating inferences on whether a rocky exoplanet has an atmosphere, how thick an atmosphere is, and what the thermal structure of the atmosphere might be.

The first major effect of thermally emitting and reflecting surfaces, as described in \citet{Hammond2025}, is that rocky exoplanets without atmospheres can have spectral features (that can potentially mimic atmospheric absorption) due to a wavelength-dependent surface albedo. This is highlighted in the top, left panel of Figure \ref{fig:main-takeaways} which displays the secondary eclipse of a nominal rocky exoplanet orbiting an M-dwarf star\footnote{Properties taken from TOI-1685b: R$_p$ = 1.468 R$_\mathrm{E}$, M$_p$ = 3.03 M$_\mathrm{E}$, a$_p$ = 0.01138 AU, R$_\star$ = 0.4555 R$_\odot$, T$_\star$ =3575K}, assuming no atmosphere and different surface compositions (ultramafic and granitoid from \citealt{Hu2012} and basaltic (tholeiitic basalt) from \citealt{Hammond2025}) vs a gray blackbody (albedo = 0) surface. Each lab-data surface imparts compositional-specific features in infrared (1-15 \textmu m) wavelengths, and reflect differing levels of light in shorter wavelengths (0.30-1 \textmu m). 

The second major effect is that there is a tradeoff between features of atmospheric origin and surface origin that depends on the thickness of the atmosphere (corresponding to the surface pressure parameter). This is displayed in the top, right panel of Figure \ref{fig:main-takeaways}, where we model a 100\% CO$_2$ atmospheres with a gray blackbody and granitoid surface. Rocky planets with \nt{thicker} atmospheres (1 bar, red) will have spectra dominated by gas absorption and emission, whereas rocky planets with thin atmospheres (1e-4 bar, orange) will have spectra dominated by surface features. Intermediate pressures will have both atmospheric and surface features (1e-2 bar, dark green in bottom, left panel). The inclusion of surfaces with wavelength-dependent albedo in a retrieval framework allows for more accurate retrieved surface pressure values by fitting for the relative strengths of atmospheric and surface features. 

The third major effect is that surfaces can impart pseudo-features in isothermal atmospheres due to transparency windows. In emission, non-isothermal pressure-temperature profiles are derived by fitting absorption features (indicative of a decrease in temperature with increasing height/decreasing pressure) or emission features (indicative of thermal inversions). Isothermal atmospheres, or regions of the atmosphere that are isothermal, produce no observable features in the spectrum and only contribute blackbody flux. However, due to surfaces having wavelength dependent albedo, they can impart pseudo-features in wavelengths where there are gas-opacity transparency windows. Displayed in the bottom, left panel of Figure \ref{fig:main-takeaways}, we model an atmosphere with a pressure-temperature gradient (green) and see both CO$_2$ and surface features in the spectra. The atmosphere with an isothermal pressure-temperature profile (turquoise) produces a featureless, blackbody emission spectrum in wavelength regions where CO$_2$ has strong absorption, whereas where CO$_2$ has no strong absorption, the spectrum follows the surface albedo spectrum (analogous to `transparency windows' on Earth). When a gray blackbody surface is used, this effect does not show up in the modeled spectra. 

Finally, surfaces can flip absorption features into emission features even if the pressure-temperature profile does not have an inversion. In the bottom, right panel of Figure \ref{fig:main-takeaways} we model an 100\% N$_2$ atmosphere (which has near negligible absorption features), sub-micron hematite clouds (Fe$_2$O$_3$), and a pressure-temperature gradient. At high surface pressures (\nt{thicker} atmosphere, blue), any effect from the surface is not present in the spectra and the hematite \nt{aerosols} produces an absorption feature.  Hematite has an absorption feature in a wavelength region where granitoid has a high albedo ($\sim$8.5 \textmu m), and therefore a lower effective surface temperature (see \S \ref{sec:2.1}). At lower surface pressures (purple), this lower effective surface temperature compared to the atmospheric layers above it causes a surface-atmosphere interface pseudo-temperature inversion that flips the hematite absorption feature into emission. \nt{In other words, the specific intensity from the surface, $\epsilon_{\mathrm{surface}}$$B(T_{\mathrm{surface}})$, is lower than the source function of the atmosphere, $B(T_{\mathrm{atmosphere}})$, at wavelengths where the surface emissivity is low.} Without thermally emitting and reflecting surfaces, this effect would be attributed to soley an atmospheric temperature inversion. 

The inclusion of surfaces with wavelength-dependent albedo into a retrieval framework can help ensure that any inference made about the atmosphere from an observed spectra (its existence, the thickness, and the pressure-temperature profile) are more robust. \nt{Degeneracies in solution space, between forward models, have already been highlighted from both spectral \citep[e.g.,][]{Luque2024} and photometric \citep[e.g.,][]{Holmberg2026} datasets.} Retrieval frameworks with surfaces \nt{and overlaying atmospheres} can capture degenerate solutions between, for example, an emission feature caused by an atmospheric inversion versus a thin atmosphere with a high-albedo surface, \nt{with a single retrieval model}.

In the next section, we explore how the intricacies of jointly considering surfaces and atmospheres in a retrieval framework helps to constrain the surface pressure. 

\section{Retrieval Analysis of TOI-1685b and 55 Cancri e: Thin vs Thick Atmosphere}\label{sec:4}

\begin{figure*}[ht!]
    % \vspace{-1.3cm}
    \includegraphics[width=1.0\textwidth]{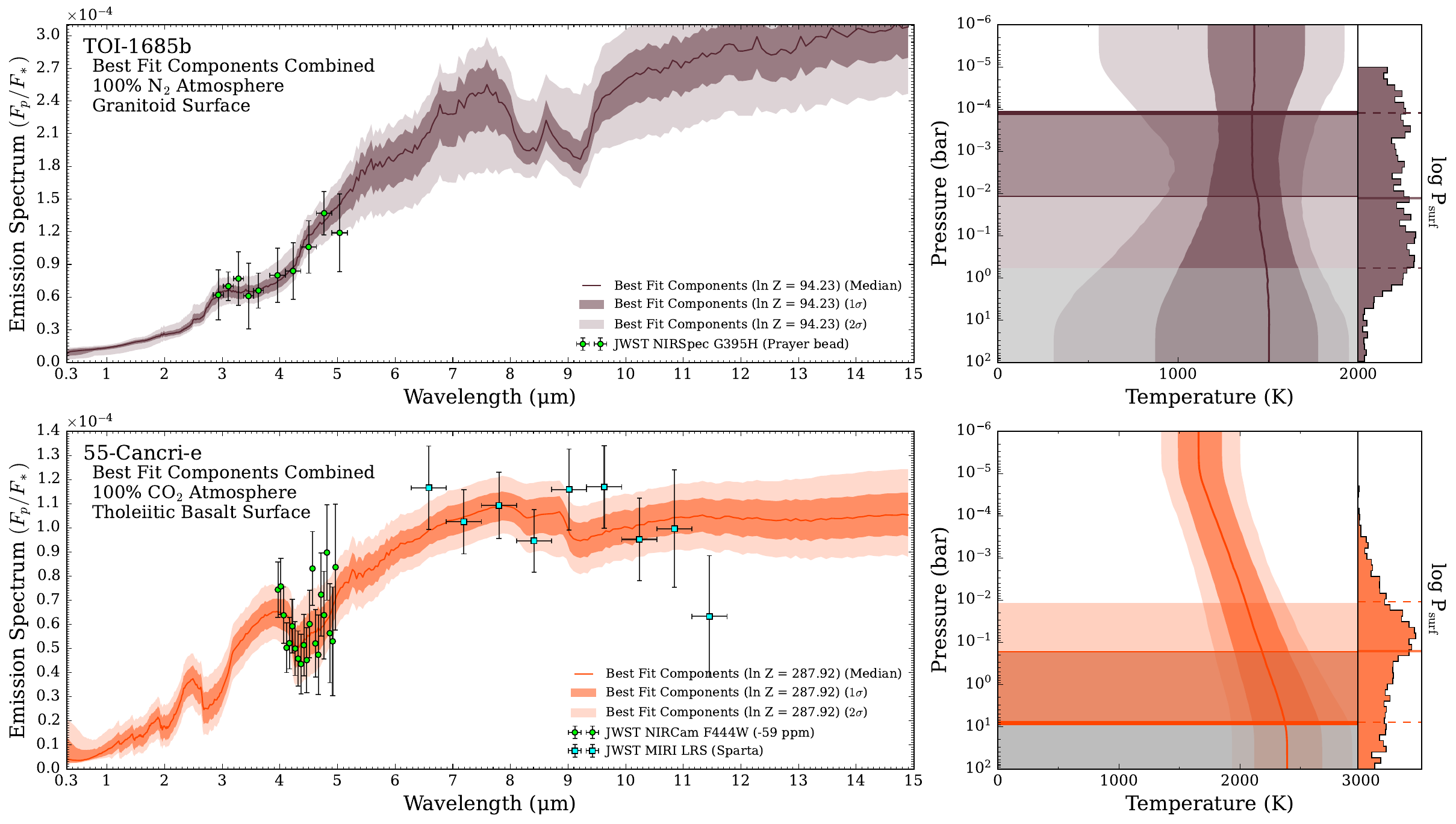}
    \caption{Retrievals of TOI-1685 b and 55 Cancri e utilizing the highest evidence components from a suite of simplified retrievals (as described in \S \ref{sec:4}). These retrievals include an emitting and reflecting surface, parameterized by the surface pressure log P$_{\mathrm{surf}}$, and overlaying atmosphere composed of a single gas, and patchy Mie-scattering clouds. Panels display 2$\sigma$ retrieved spectra and pressure-temperature profile, as well as an inset displaying the posterior distribution for the retrieved surface pressure. \textbf{Top} We find that retrieval analysis of TOI-1685 b's NIRSPEC G395H data results in a low surface pressure ($\leq$ 1 bar), agreeing with previous studies \citep{Luque2024} that the object is most likely a bare rocky planet. \textbf{Bottom} Retrieval analysis of 55 Cancri e's NIRCam and MIRI LRS data results in a higher surface pressure ($\geq$ \nt{0.1 bar}), agreeing with previous studies \citep{Hu2024} that the object has a thick, outgassed atmosphere. This constraint is due to the CO$_2$ absorption feature in the NIRCam data. \nt{The retrieved surface pressure posteriors are a result of the impact surfaces have on observed spectra (as described in Figure \ref{fig:main-takeaways}), allowing JWST quality data to distinguish between thin and thick atmospheres} (see Figure \ref{fig:turning-knobs} for a visual exploration of how surface pressure was constrained). For our one-component retrieval results, and the version of this figure with the full posteriors, see Figures 1-7 in Zenodo Supplementary Material (see \S \ref{sec:zenodo}).}
    \label{fig:retrieval-psurf}
\end{figure*}

We test our new \texttt{POSEIDON} features by performing retrieval analyses on the publicly available secondary-eclipse data of TOI-1685b \citep{Luque2024} and 55 Cnc e \citep{Hu2024}, representing objects with evidence of a thin and thick atmosphere, respectively. 

TOI-1685b is a super-Earth (R$_\mathrm{p}$ = 1.468 R$_\mathrm{E}$, M$_\mathrm{p}$ = 3.03 M$_\mathrm{E}$, a$_\mathrm{p}$ = 0.01138 AU) orbiting an inactive M-dwarf star with an equilibrium temperature of $\sim$1000K. \citet{Luque2024} observed a JWST NIRPsec G395H (3.823-5.172 \textmu m) phase curve of this object and found that the emission spectrum is best fit by a featureless blackbody with no heat redistribution (i.e., no thick atmosphere) and a low albedo (\nt{where the observed dayside brightness temperature is near equal to the dayside brightness temperature of a zero Bond albedo, zero heat recirculation blackbody, $\sim$1390K}). For our retrievals, we use the `prayer-bead' dataset, but utilize averaged error bars due to \texttt{POSEIDON} not having current support for asymmetric error bars. 

55 Cancri e is a super-Earth (R$_\mathrm{p}$ = 1.95 R$_\mathrm{E}$, M$_\mathrm{p}$ = 8.8 M$_\mathrm{E}$, a$_\mathrm{p}$ = 0.01544 AU) orbiting a bright K-dwarf with an equilibrium temperature of $\sim$2000K, which is hot enough to vaporize rock. \citet{Hu2024} observed a JWST NIRCAM F44W (3.94-4.49 \textmu m) and MIRI LRS (5-12 \textmu m) secondary eclipse of the object and found that the spectrum could be best fit by a thick, outgassed CO$_2$ or CO atmosphere, ruling out a thin atmosphere made of vaporized rock \citep[e.g.,][]{Schaefer2009,Miguel2011,Ito2015,Kite2016}. For our retrievals, we use both datasets and allow for a free-offset between them, as was done in \citet{Hu2024}. 

With our goal to run a retrieval including an emitting and reflective surface, an overlaying atmosphere, and patchy clouds we first run a suite of simplified retrievals to guide our model parameterization choice. Specifically, we first run retrievals fitting the datasets assuming the atmosphere is 100\% one gaseous species while also fitting for a gray, blackbody (albedo = 0) surface \nt{pressure} and a gradient pressure-temperature profile (defined by two parameters, T$_{\mathrm{high}}$ and T$_{\mathrm{deep}}$). For TOI-1685b we test atmospheric compositions of CH$_4$ \citep{Yurchenko2024}, CO \citep{Li2015}, CO$_2$ \citep{Yurchenko2020}, H$_2$ \citep{Hohm1994,Karman2019}, H$_2$O \citep{Polyansky2018}, N$_2$ \citep{Karman2019}, O$_2$ \citep{Gordon2022}, and SO$_2$ \citep{Underwood2016} and find that N$_2$ has the highest evidence. We then run retrievals with Mie-scattering aerosols (Exohaze (300K) \citep{He2024}, Fe$_2$O$_3$ \citep{Triaud2005}, H$_2$O (liquid) \citep{Hale1973}, H$_2$O (ice) \citep{Warren1984}, H$_2$SO$_4$ \citep{Palmer1975}, S$_8$ \citep{Palik1998}, SiO \citep{Palik1985}, and Tholins \citep{Khare1984,Ramirez2002}) and find that SiO has the highest evidence. \nt{For simplicity, our suite of aerosol exploratory retrievals utilize a non-patchy model where aerosols are described by a mean particle radius (log $r_\mathrm{m}$, with an assumed lognormal radius distribution, width = 0.5, centered on a mean radii) and a constant-with-pressure volume mixing ratio (i.e., `Uniform-X' in \citealt{Mullens2024}). For our combined model, we utilize the same model but with an additional $f_\mathrm{cloud}$ parameter that describes the percent of aerosol coverage (where resultant spectra are a weighted sum of emergent spectra from a clear and a cloudy sector).}

Finally, we perform bare-rock retrievals with all three surface models introduced in \S \ref{sec:2} and find that a granitoid surface \citep{Hu2012} has the highest evidence. All single-component retrievals can be found in Figures 1-3 in Zenodo Supplementary Material (see \S \ref{sec:zenodo}). We then perform a retrieval with the highest evidence components integrated together into a single model: a 100\% N$_2$ atmosphere, granitoid surface, and patchy SiO clouds. 

We perform a similar retrieval analysis for 55 Cancri e. For the atmospheric composition retrievals we find that a 100\% CH$_4$ atmosphere had the highest evidence, but we instead choose to use the second highest-evidence retrieval with 100\% CO$_2$ since the CH$_4$ retrieval has an nonphysical pressure-temperature profile inversion and forward models do not predict large abundances of CH$_4$ at the high temperatures of 55 Cancri e \citep{Moses2013}. We test a myriad of Mie-scattering mineral aerosols (due to the prediction that rocks vaporize on the surface of 55 Cancri e; Al$_2$O$_3$ \citep{Koike1995}, Meteoritic Diamonds \citep{Mutschke2002}, Fe$_2$O$_3$, Fe$_2$SiO$_4$ \citep{Fabian2001}, MgSiO$_3$ (amorphous) \citep{Egan1975,Dorschner1995}, Mg$_2$SiO$_4$ (crystalline) \citep{Suto2006}, SiO, and SiO$_2$ (crystalline) \citep{Palik1985,Zeidler2013}, and find that Meteoritic Nano-Diamonds has the highest evidence. Finally, our bare-rock retrievals find that find that a basaltic (tholeiitic basalt) surface \citep{Hammond2025} has the highest evidence. We then perform a retrieval with the highest evidence components integrated together into a single model: a 100\% CO$_2$ atmosphere, tholeiitic basalt surface, and patchy diamond clouds. All single-component retrievals can be found in Figures 1-6 in Zenodo Supplementary Material (\S \ref{sec:zenodo}, see Figures 3 and 6 in particular to see which specific lab albedos were tested).

\nt{Our single-component retrievals, which guide our definition of a maximally complex model, are representative how analyses of JWST rocky planet emission spectra has been done in previous works \citep[e.g.,][]{Luque2024,Hu2024}, where fits are accomplished by assuming either a bare rocky surface or a thick atmosphere (Figure \ref{fig:first-figure}). Our goal in this work is to allow for a fully data-driven inverse fit of the observed spectra that allows for both thin- and thick- atmospheric solutions to see if the final result matches previous claims. We note that our combined retrieval model represents only a subset of possible model configurations, and that a thorough retrieval exploration that includes multiple surface-atmospheric parameters simultaneously will provide more robust results that don't exclude plausible global solutions, but is outside the scope of this work.}

The reference pressure was set at 10$^{-2}$ \,bar with the model pressure grid covering 10$^{-6}$ - 100 bar with 100 layers uniformly distributed in log-pressure space. We take the surface pressure prior to range from 10$^{-5}$ - 100 bar due to instabilities in the thermal scattering and reflection modules when the number of atmospheric pressure layers above the surface reaches unity. Our \texttt{POSEIDON} retrievals sample the parameter space using \texttt{MultiNest} \citep{Feroz2008}, here with 1,000 live points. Model emission spectra are computed at a spectral resolution of R = 10,000 in regions of wavelength space with data coverage, and R = 1000 for regions without data coverage (where gas opacities are pre-loaded line-by-line at this resolution). All retrievals with atmospheres utilized the thermal multiple scattering and reflection radiative transfer introduced in \citet{Mullens2024}, and all retrievals with aerosols utilize the pre-computed database with updated scattering properties described in \citet{Mullens2024} and \citet{Mullens2025}. Retrieved 2$\sigma$ spectra, alongside the retrieved posteriors for surface pressure, can be seen in Figure \ref{fig:retrieval-psurf}. A companion figure with the full posteriors can be seen in Figure 7 in Zenodo Supplementary Material (see \S \ref{sec:zenodo}).

We find that, in agreement with previous studies, the TOI-1685b dataset prefers a thin atmosphere (log P$_{\mathrm{surf}}$ $\leq$ 1) with an \nt{unconstrained,} near-isothermal \nt{median} pressure-temperature profile while the 55 Cancri e dataset prefers a \nt{thicker} atmosphere (log P$_{\mathrm{surf}}$ $\geq$ -1) with a gradient pressure-temperature profile. In both retrievals, Mie-scattering aerosol properties are not well-constrained (Figure 7 in Zenodo Supplementary Material). \nt{Our method of allowing a fully data-driven inverse method to choose its preferred solution aligns with previous studies while allowing for additional insights.}

\nt{For TOI-1685b, \citet{Luque2024} found that forward-model atmospheric solutions resulted in data-consistent fits for surface pressures ranging from 1 bar (for 100\% H$_2$O) to 1e-4 (for 100\% CH$_4$). Their bare rocky surface models using a blackbody substellar temperature of $\sim$1550K find that all surface models with low-albedo datasets (e.g., basaltic) are consistent. Our retrieval, which allows for both thick and thin atmospheric solutions, retrieved a pressure-temperature profile centered on $\sim$1500K (while displaying the full range of potential solutions for surface temperature in the 2$\sigma$ solution) and fits for a range of permissible surface pressures that have an upper limit of $\sim$1 bar. For 55 Cancri e, in \citet{Hu2024} they note that a high albedo surface with no atmosphere could mimic the observed dataset, which they reject based on a physical argument (i.e., molten silicates have low albedos). They also perform atmospheric retrievals that retrieve surface pressures of 1e-2 to 100 bar. We find that our retrieval returns a similar surface pressure range while also placing the surface (which has a lab-derived albedo) at higher pressures.}

By performing retrievals with emitting and reflecting surfaces, overlaying atmospheres, and patchy Mie scattering clouds we can statistically assert that JWST quality data can distinguish between thin and thick atmospheres \nt{while gaining insights into how a specific solution was preferred}. In particular, we find that the effects of including an emitting and reflecting surface in our radiative transfer, featured in Figure \ref{fig:main-takeaways}, helps constrain the surface pressure. As shown in Figure \ref{fig:turning-knobs}, even though the retrieved pressure-temperature profile of TOI-1685b is near-isothermal, due to the N$_2$-N$_2$ opacity in the 3-5 \textmu m region there are pseudo-features in the spectrum when the atmosphere is thick (similar to the effect showcased in the bottom left panel of Figure \ref{fig:main-takeaways}). In the case of 55 Cancri e, as shown in Figure \ref{fig:turning-knobs}, low surface pressures can cause a surface-atmosphere \nt{pseudo-}temperature inversion and flip the CO$_2$ feature from absorption to emission (similar to the effect showcased in the bottom right panel of Figure \ref{fig:main-takeaways}). 

\nt{We note here, as an aside, that there are disputing claims on the existence of an atmosphere on 55 Cancri e. Our retrieval prefers the existence of an atmosphere due to the CO$_2$ absorption feature in the NIRCam emission data. Other work \citep{Patel2024} shows that NIRCam eclipse data is highly variable visit-to-visit (either due to difficulties in data reduction from such a bright target or atmospheric variability), while HST transmission data \citep{Tsiaras2016} claim an H$_2$ atmosphere with trace amounts of HCN. In this work we follow the results of \citet{Hu2024}, but we note that future JWST Cycle 5 observations (JWST-GO-9825, PI: Hoeijmakers \& JWST-GO-12237, PI: Boehm)  are slated to reobserve 55 Cancri e and determine whether the variability is systematic or due to an atmosphere.}

The inclusion of surfaces with overlaying atmospheres in radiative transfer can help distinguish between thick and thin atmospheres in a retrieval framework. Now, we explore JWST's potential to constrain the surface composition of bare rocky exoplanets. 

\section{Detecting Surface Geology with JWST}\label{sec:5}

\begin{figure*}[ht!]
\centering
    % \vspace{-1.3cm}
    \includegraphics[width=1\textwidth]{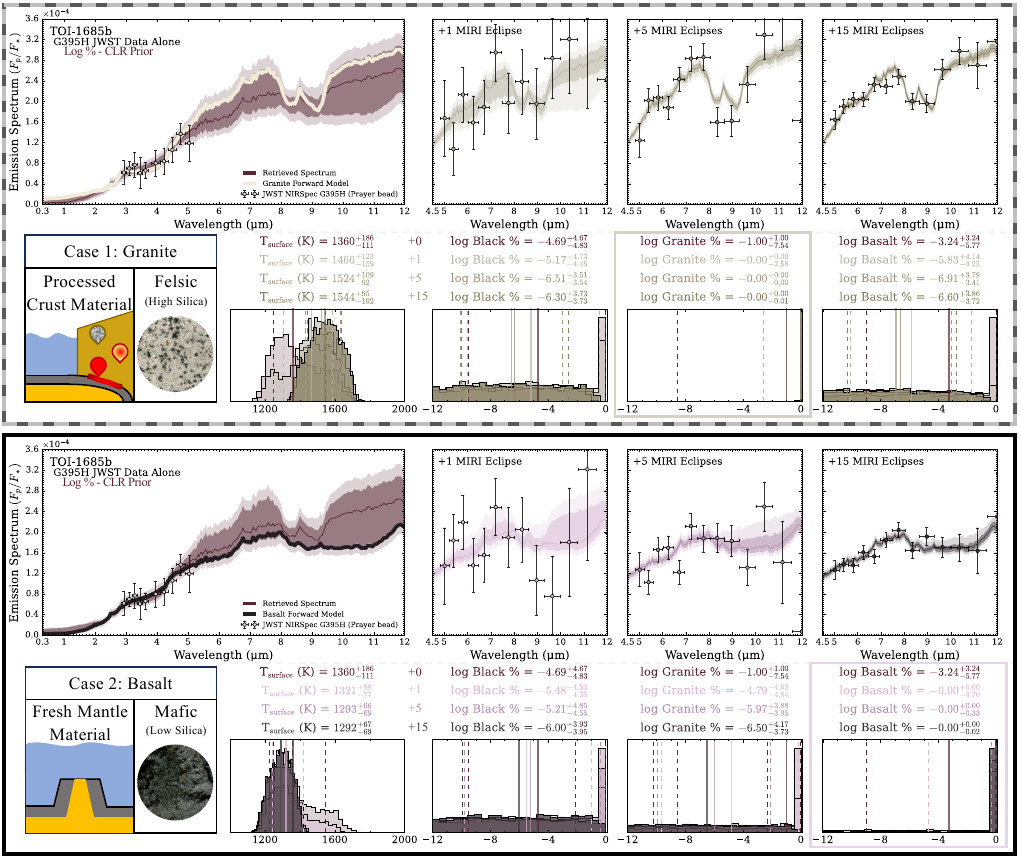}
    \caption{With JWST MIRI LRS, surface geology of thin-atmosphere (or airless) rocky exoplanets can be probed by detecting diagnostic mid-infrared features. We perform bare-rock, multi-surface (`black' being a constant albedo of 0, granitoid \citep{Hu2012}, and tholeiitic basalt \citep{Hammond2025}) component retrievals with the CLR prior \nt{and log surface component percentages} on current and simulated (with \texttt{PANDEXO}, \citealt{batalha2017_pandexo}) TOI-1685b datasets. \textbf{Top} Retrieval results for granite, which represents a surface geology composed of an silica-rich (felsic) rock. On Earth, granitic continental crust is generated from the partial melting of oceanic crust (where the melt is silica-rich) and is a direct result of plate tectonics, however, on other worlds granite can arise without plate tectonics via exotic mantle compositions (see Table \ref{table:Surface Albedos Categories 2}). Top panels display results of multi-surface retrievals on current NIR data, +1, +5, and \nt{+15} simulated MIRI eclipses (R = 15). Bottom panels display resultant histograms, with a granite surface confidently detected as the \nt{sole} surface component by 5 eclipses. \textbf{Bottom} Retrieval results for basalt, which represents a surface geology composed of an relatively silica-poor (mafic) rock. On Earth, basaltic oceanic crust is generated from the cooling of material originating from the mantle. Basalts are the most common surface rock of Solar System terrestrial worlds, a pattern expected to extend to exoplanets (see Table \ref{table:Surface Albedos Categories 2}). A basaltic surface is confidently detected as the \nt{sole} surface component by 15 eclipses. We note that the SNR modeled in our test case is unique to TOI-1685 b, and that other targets might need more or less eclipses to detect surface geology. See \nt{Figure \ref{fig:Linear-JWST-sims} for a version of these retrievals that instead use a uniform prior with linear surface component percentages}, and see Figure 8 in Zenodo Supplementary Material (see \S \ref{sec:zenodo}) for a uniform prior with log surface component percentages. (Granite and Basalt pictures sourced from the \href{https://naturalhistory.si.edu/explore/gems-and-minerals/geogallery}{Smithsonian NMNH GeoGallery}.)}    
    \label{fig:CLR-JWST-sims}
\end{figure*}

As discussed in \S \ref{sec:3}, bare rocky surfaces with no or very-thin atmospheres can have spectra with surface-identifying features due to surfaces having wavelength-dependent albedos, often in mid-infared wavelengths accessible with JWST MIRI LRS (5-12 \textmu m, see Figure \ref{fig:main-takeaways} top left panel \nt{and also \citealt{Luque2024}}). In this section, we aim to determine whether or not \nt{simulated} JWST quality data can distinguish between different surface geologies. 

We first set up a multi-component bare rocky surface retrieval with the inclusion of granitoid  \citep{Hu2012} and basaltic \citep[specifically tholeiitic basalt,][]{Hammond2025} \nt{surface components}, as well as `black' (which is just a surface with a constant albedo of 0) \nt{for} the JWST NIRSpec G395H secondary-eclipse data of TOI-1685b \citep{Luque2024}. We chose TOI-1685b since the analysis done in \citet{Luque2024} and this work (\S \ref{sec:4}) point towards it being a bare-rock, and (at the time of this work) there are currently no secondary-eclipse JWST MIRI LRS datasets. 

We perform our retrievals by testing \nt{a uniform prior (testing both linear and log surface percentage components), and a CLR prior (with only log surface percentage components, since the CLR prior requires log).} \nt{We set} the `black' surface in the CLR retrievals as being the non-free parameter, where its percentage is computed by being the percentage needed to add to a total of 100\% (see \S \ref{sec:2} for details). We find that with the current JWST near-infrared dataset, the retrievals cannot distinguish between blackbody, granite-like, and basaltic surfaces, \nt{in line with the results of \citet{Luque2024}} (Figures \ref{fig:CLR-JWST-sims} and \ref{fig:Linear-JWST-sims}, left panels).

Specific surfaces are direct indicators of crustal composition and evolution. On Earth, both granites and basalts are a major component of the crustal surface; continental crust is granite-rich and oceanic crust is basalt-rich. Granite is a silica-rich (i.e., felsic; $\geq$ 63 wt\% SiO$_2$) rock that is generated \nt{primarily} from the partial melting of oceanic crust at oceanic-continental subduction zones, where the portion of the crust that melts is more silica-rich and less dense than the parent material, where it then rises and cools at the surface. In the Solar System, only Earth has a large granitic component to its surface, which has been attributed to its unique geologic process of plate tectonics. However, granitic surfaces could potentially form on exoplanets without these processes; for example, if a rocky planet has an initial composition that is much more silica-rich, as evidenced by the exotic mantle compositions gleaned from white dwarf pollution observations \citep{Putrika2021}. Here we specifically use a modeled granite-like surface \citep{Hu2012} that combined individual mineral albedos with the Hapke radiative transfer method \citep{Hapke1981,Hapke2002} (here, 40\% K-feldspar, 35\% quartz, 20\% plagioclase, 5\% biotite) where the strong double-peaked feature in the mid-infrared (8-9 \textmu m) can be attributed to mainly quartz. 

Basalt is a silica-poor (i.e., mafic; $<$ 63 wt\% SiO$_2$) rock that is the natural consequence of partially melted mantle material cooling. Large amounts of fresh basalt is formed on Earth via the cooling of mantle material at mid-ocean ridges, and most Solar System planets have a large component of their observable crust composed of basalts, a trend expected to extend to rocky exoplanets \citep{Putirka2019,Fortin2022,First2025}. Regardless, basalts on the surface generally indicate that a planet has experienced differentiation and the formation of a secondary crust (see Table \ref{table:Geology Glossary}). Here we specifically use albedo data of a natural thoeletiic basalt sample (`Taos Basalt slab') \citep{Hammond2025}, which is a more silica-rich form of basalt produced at mid-ocean ridges on Earth. The broad mid-infrared (8-12 \textmu m) absorption feature can be found in many basaltic samples \citep[e.g.,][]{First2025} and can't be attributed to a single mineral due to basalts not being fully crystallized \citep{Putirka2021}. 
%In Tables \ref{table:Surface Albedos Categories} and \ref{table:Surface Albedos Categories 2}, we have provided Solar System contexts for common geologic categories as well as potential considerations for interpretation if a specific surface geology is detected on an exoplanet. 

% In particular, granites are indicative of a surface formed from intrusive igenous rocks (formed from magma) with high ($\geq$ 63\%) SiO$_2$ content (felsic) that are brought to the surface via processes like plate tectonics, while basalts are indicative of a surface formed from extrusive igenous rocks with low ($\leq$ 45\%) SiO$_2$ content (mafic) that form on the surface via the cooling of lava (and are predicted to be the most common surface for terrestrial worlds, e.g., \citet{First2025}).

Using these two albedo datasets as a basis, we then simulated 1, 5, and \nt{15} JWST MIRI LRS (R = 15) eclipses \nt{(where 1 and 5 eclipses represent a reasonable amount of visits in a single JWST GO program, while 15 is idealized)} from spectra generated assuming 100\% of the surface is granitoid or basalt with Pandexo \citep{batalha2017_pandexo} and ran retrievals on the joint datasets. \nt{Results utilizing a CLR prior with log percentages can be seen in Figure \ref{fig:CLR-JWST-sims}, a uniform prior with linear percentages in Figure \ref{fig:Linear-JWST-sims}, and a uniform prior with log percentages in Figure 8 in the Zenodo Supplementary Material (see \S \ref{sec:zenodo}).} We find that, \nt{for this specific test case} using \nt{simulated} JWST MIRI LRS data \nt{and idealized forward models}, retrievals utilizing the CLR prior \nt{differentiates between the surface being solely} granitoid and a basaltic \nt{surface within $\sim$5 and 15 \nt{eclipses} (respectively).}

\nt{We note there are key differences and strengths to using the CLR prior with log surface component percentages versus a uniform prior with linear surface component percentages. Surface components that are less than order-unity percentages have a minimal impact on a resultant spectra \citep[e.g.,][]{Zelakiewicz2026}. This differs from gas and aerosol opacities, which at trace amounts (volume mixing ratios $\sim$ -10 to -1) can have absorption features that vary by order of magnitudes at specific wavelengths, and therefore imprint on observed spectra \citep[e.g.,][]{Grant2023,Louie2025}. The strength of a CLR prior with log percentages is that it is optimized to weigh how likely a surface component is to contribute the most to an observed spectrum (i.e., the `bulk', or sole surface component), which can be seen in the histograms of Figure \ref{fig:CLR-JWST-sims}. On the other hand, a uniform prior with linear percentages is more adept at capturing potential degeneracies that can occur from linearly combining spectra from separate surface component albedo datasets \citep{Zelakiewicz2026}. These unity-order surface percentage degeneracies can be seen in Figure \ref{fig:Linear-JWST-sims}. For the uniform prior with linear percentages, at 5 eclipses granite can be detected (where the retrieved median at 5 eclipses is closer to the modeled true value of 100\% granite than at 15 eclipses because of a noise draw that placed the granite feature deeper than its true value). Basalt has a more difficult time fitting up to 100\%, even at 15 eclipses, though it still remains more constrained to higher percentages than granite. This result is probably due to a lack of the necessary SNR at 10-11 \textmu m where the basalt feature is, and/or the shape of the JWST NIRSpec G395H dataset, to favor solely basalt over granite in these retrievals.}

% We do note, however, that the ultramafic dataset in particular is still somewhat consistent with a blackbody surface due to its surface feature being present in lower SNR wavelengths (10-12 \textmu m), and would probably require additional eclipses to definitely detect. 

This work displays the \nt{potential} power of JWST to detect surface geologies of rocky bodies with thin or no atmospheres, opening the door for comparative geology between rocky bodies in our Solar System and rocky bodies in other systems. \nt{Observing programs}, such as JWST-GO-7953 (PI: Dr. Kimberly Paragas) will be leveraging JWST to begin this exciting work; however, there remains many unknowns when predicting and interpreting the surface geology of rocky exoplanets\footnote{For a tutorial on how to perform thermal emission retrievals for bare rocky exoplanets in \texttt{POSEIDON}, see the \nt{`Thermal Emission Retrievals of Bare Rocky Exoplanets;} tutorial in \href{https://poseidon-retrievals.readthedocs.io/en/latest/content/retrieval_tutorials.html}{Retrieval Tutorials}.}. 

\subsection{Predicting and Interpreting Surface Geology}\label{sec:5.1}

In order to predict the geologic composition of the topmost observable layer of a planet, one must first predict the composition of the planet's mantle and core. A common method is to model the `equilibrium' bulk composition and mineralogy of the mantle and core based on the bulk Mg/Si (and occasionally the Fe/Mg ratio), where it is usually assumed that a planet's refractory element relative abundances are similar to its host star's measured abundances \citep[a pattern confirmed to be true in Solar System terrestrial planets, with the exception of Mercury, see review][]{Putirka2021} from the \texttt{Hypatia Catalog} \citep{Hinkel2014}. Pressure-dependent internal mineralogies can then be predicted with a phase-diagram analysis \citep{Hinkel2018,Hinkel2024}, which can include additional parameters such as devolitization, fractionation of elements between the core and mantle, oxygen fugacity, and an intrinsic pressure-temperature profile \citep[see \texttt{pyExoInt},][]{Wang2019,Wang2022}, or the use solar-system informed metallic iron-core formation mechanisms to determine the resultant mantle composition \citep{Putirka2019}. The \texttt{Hypatia} informed modeling from \citet{Putirka2019} and \citet{Wang2022} suggests that many exoplanets will have Earth-like mantles composed of pyroxenite and peridotite (see Table \ref{table:Surface Albedos Categories} for definitions), however, their work includes compositional outliers and recent white dwarf pollution observations have detected exotic mantle compositions \citep{Putrika2021}. 

These models not only predict the composition of minerals and resultant rocks that will dominate the mantle reservoir that can form the surface, but also have implications for the thermal properties of a planet (i.e., for tidally locked exoplanets, whether the substellar surface is expected to remain melted as a magma ocean or cool into a crust), the mantles ability to outgas or dissolve atmospheric gases \citep{Brugman2021,Bello-Arufe2025}, and many more crucial properties that are necessary to interpret observed surface and atmospheric signatures \citep[see the magma ocean-atmospheric model intercomparison project, ][]{Lichtenberg2025}.

In a generalizable sense, a planet's crust, and therefore the observable portion of the planet, will form from either the initial cooling of the primordial magma ocean, the partial melting of the mantle, or any subsequence alteration of the crust from repeated melting and interactions with an atmosphere (and for cooler objects, surface liquid water). In the case of the primordial Earth and the Moon, the first crust formed from the cooling of the magma ocean, in which dense olivines first crystallized out and sunk leaving the resultant magma enriched in calcicum and aluminum that then formed low density plagioclase feldspars (specifically anorthite) that rose to the top (see Primary Crust in Table \ref{table:Geology Glossary}). Over time, resurfacing occurred due to the partial melting of the mantle. At depth, mantle rocks begin to fractionally melt wherein this melted material, which can erupt on the surface and form the crust (see Secondary Crust in Table \ref{table:Geology Glossary}), is enriched in materials with lower melting points (usually silica). In the case of Earth-like peridotite and pyroxenite mantles, partial melting forms a magma enriched in silica and alumina that solidify to form basalts. Basalts cover over 2/3 of the Earth's surface (in the form of fresh, oceanic crust) and nearly the entirety of Mercury, Mars, the Moon, Vesta, and probably Venus \citep{Putirka2021}. The last form of crustal material is unique to Earth (in the Solar System) and forms from repeated partial melting of basaltic crust as a direct result of plate tectonics and surface liquid water, forming silica-rich magma that cools into granite (the older continental crust of Earth, see Tertiary Crust in Table \ref{table:Geology Glossary}). Any surface composition can then be altered from igneous to sedimentary or metamorphic rocks by the presence of an atmosphere (i.e., which can weather, oxidize, or reduce any surface rocks), or liquid water (which can form clays, salts, and aqueous minerals in basalts), and geologic processes that cause high temperature or high pressure processing, all of which have lasting observable signatures even if the atmosphere, liquid water, and geologic processes no longer exists when an observation is taken \citep{First2025}. If the magma ocean doesn't solidify, as predicted for the daysides of tidally-locked rocky exoplanets with temperatures exceeding $\sim$1300K (for Earth-like mantle compositions, where this number can change due to melting temperature depending on composition), the detected magma composition will be largely dictated by the amount of mixing, fractional crystallization, and atmospheric interactions \citep[see review, ][]{Chao2021}.

We note that in this work we are presenting an interpretation scheme biased towards solar-system informed processes that are calibrated to pyroxenite-peridotite mantle compositions \citep[see succinct review,][]{Zalasiewicz2016}, and the exact mechanisms by which exoplanets form their observable surface is a complex function of mantle composition, incident stellar irradiation, past-and-present atmospheric and liquid water interactions, time-dependent resurfacing and alteration processes, and size (i.e., super-Earth sized rocky exoplanets might have heat-release and resurfacing mechanisms that differ from Earth's plate tectonics, Venus's resurfacing, and Mar's volcanism \citealt{Luo2024}). There are currently many unknowns when predicting and interpreting detected surface geology (i.e., would a detected granitic surface indicate the presence of current or past Earth-like plate tectonics, the cooling of a planet that formed with an initially silica-rich mantle, or the resurfaced metamorphic transformation of rock into a mineralogical form similar to granite, e.g., gneiss?), and atmospheric signatures that can be influenced by interactions with surface compositions \citep[i.e., 55 Cancri e][]{Hu2024} or volcanic outgassing \citep[i.e., L 98-59 b][]{Bello-Arufe2025}. 

In order to begin and create a framework by which to geologically interpret the observed surface composition of an exoplanet, we have compiled a list of pertinent geology vocabulary in Table \ref{table:Geology Glossary}, and broad geological categories in Table \ref{table:Surface Albedos Categories} and \ref{table:Surface Albedos Categories 2} to sort the albedo lab data in Tables \ref{table:Surface Albedos}, \ref{table:Basalt-Library}, \ref{table:Lava-Library} and relate geology to exoplanet science. We note that our presented classification schemes are just skimming the surface of information available to be extracted from the field of geology, and only cover pertinent categories for our curated albedo database. We hope that these supplementary tables facilitate future inter-disciplinary collaborations with geologists and astronomers as the field of exogeology grows and evolves.

\section{Discussion and Summary}\label{sec:last}

In this work, we introduced new features to the retrieval code \texttt{POSEIDON} to include thermally emitting and reflecting surfaces in its radiative transfer for both atmospheres (clear and cloudy) and bare rocky surfaces. We find that the inclusion of surfaces with lab-derived albedos strongly affects the emergent spectra of rocky exoplanets: 

\begin{itemize}
\item Surfaces modeled with geologically-specific albedos impart mid-infrared absorption features that can mimic atmospheric absorption.
\item Atmospheric and surface-origin absorption features can trade-off, depending on a planet's surface pressure.
\item Surfaces can impart pseudo-features due to atmospheric transparency windows (even for atmospheres with isothermal pressure-temperature profiles).
\item Surface-atmospheric interfaces can cause pseudo-temperature inversions that flip absorption features to emission.
\end{itemize}

%To summarize, surfaces impart mid-infrared absorption features that can mimic atmospheric absorption, there is a trade-off between the strength of absorption features of atmospheric and surface-origin depend on a planet's surface pressure, surfaces can impart pseudo-features due to atmospheric transparency windows (even in atmospheres with isothermal pressure-temperature profiles), and surface-atmospheric interfaces can cause inversions that flip absorption features to emission. 

Including these important spectral effects arising from surfaces allows retrieval frameworks to both constrain surface pressures and capture degeneracies between pressure-temperature profiles and surface properties. We have showed that, taken together, these spectral signatures are a powerful diagnostic of surface pressure that can distinguish between thick and thin atmospheres in already-available JWST emission spectra. We explored JWST MIRI LRS's capability to constrain surface geology of exoplanets, by detecting specific mid-infrared surface signatures, and showed that JWST is well suited to detect mid-infrared features from specific surface compositions. 

Our update to \texttt{POSEIDON} includes an open-source database of lab surface data tailored for exoplanet-specific science, standardized when possible to a specific lab-measurement (directional-hemispherical reflectance). In order to contextualize our geology databases (Tables \ref{table:Surface Albedos}, \ref{table:Basalt-Library}, \ref{table:Lava-Library}) and foster collaboration between geologists and exoplanet scientists, we developed three supplementary tables (Tables \ref{table:Geology Glossary}, \ref{table:Surface Albedos Categories}, \ref{table:Surface Albedos Categories 2}). Taken together, \texttt{POSEIDON} will be well suited for emission and reflection retrievals of rocky worlds with JWST and future observatories (e.g., HWO). 

% In this work, we provide a database of surface albedos collected from libraries tailored for exoplanet-specific science, standardized when possible to a specific lab-measurement (directional-hemispherical reflectance). In order to elucidate surface geology, we developed an in-depth table sorting the lab data by geologic categories, where each category can be linked to what a detection of that mineral or rock would `signify' geologically. To account for the widely diverse nature of exoplanet science that can often defy commonly held assumptions, we have included minerals that are both common and uncommon in Solar System body surfaces, notating what processes could be at work to generate bulk surfaces on exoplanets composed of uncommon Solar System minerals. 

% Future work should involve collaboration with laboratories and geologists to create a solid link between surfaces that can be observed on exoplanets with current and future space-based observatories, and what that surface informs us geologically about the body. and what a surface informs us geologically about the body.

\subsection{Future Considerations}\label{sec:6.1}

Future analysis of JWST observations would benefit from the extant methodology developed by the field of geology. In particular, from methods developed to understand solar-system emission spectra of airless bodies. The radiative transfer of airless rocky bodies presented in this work can be extended to account for many secondary effects, including: regolith textures \citep{Paragas2025}; particle size \citep{Shirley2019}; porosity \citep{Martin2022}; thermal beaming and opposition surge \citep{Tenthoff2024,Gkouvelis2025one}; crater shadowing \citep{CAMBIANICA2024}; multiple-scattering (which is present in powdered textures and results in differing transparency features that can be indicative of potential impact bombardment \citealt{Mishra2025, Paragas2025}); space-weathering \citep[\nt{which darkens albedo and adds nanophase iron, }][]{Shirley2023,Lyu2024,Coy2025, Holmberg2026}; latitude-longitude-dependent surface temperatures and resultant flux contributions \citep{Hu2012,Hammond2025}, \nt{latitude-longitude-dependent atmospheric profiles with varying surface pressure \citep{vanBuchem2026},} and depth-dependent temperature profiles \citep{Lyu2025}. Our models could also be extended to phase-dependent observations, instead of just modeling the planet at opposition (secondary eclipse) \citep[e.g.][]{Heng2021}. While not explored here, one can in principle leverage JWST-quality data to identify the mid-infrared Christiansen Feature, a peak in emissivity before the Si-O feature, that can help determine SiO$_2$ wt\% \citep{Paragas2025,First2025}.

Our albedo database can also be extended in several ways. One addition would be to include temperature-dependent emissivity measurements, since increased temperature can change the mid-infrared absorption feature of a rock in a non-linear way due to polymorph transformations and amorphization \citep{Thompson2021,Fortin2024}. Another addition would be a suite of mineral-specific spectra that can be combined in a mixture, ensuring that observations constrain surface geologic compositions instead of fitting the absorption spectra of a specific lab-measured rock that may or may not be indicative of the chemical and temperature regime of an exoplanet surface \citep{VAUGHAN2005}. However, while mixing minerals have been shown to successfully model certain rock spectra within a margin of error (e.g., granites), in some rocks (e.g., basalts) the mixing of minerals occurs non-linearly wherein the mixing of minerals cannot match measured spectra \citep{Hu2012,Ehlmann2012}. 

\nt{While our code is well-equipped to model spectral contributions from suspended mineral aerosols, rock vapor, and volatile-rich atmospheres (\S \ref{sec:4}), as well as include surface albedo data relevant for lava worlds (Table \ref{table:Lava-Library}), we note that any interpretation of JWST spectra of hot rocky exoplanets is limited by currently available lab data of gas opacities, aerosol optical properties, and surface analog measurements, which should all be explored.}

While there are a myriad of effects one can consider, future work should explore which are directly observable with the data quality observatories like JWST provide. With JWST and future observatories, we are beginning to probe the surfaces of exoplanets. Through comparison between exoplanets and the rocky bodies in our own Solar System, JWST offers the promise of revealing the full diversity of surface and atmospheric composition on rocky worlds.

\subsection{Links and Open-Source Tutorials}\label{sec:zenodo}

The Zenodo repository for this paper (Zenodo doi: \href{https://doi.org/10.5281/zenodo.19651521}{10.5281/zenodo.19651521}) contains the `Zenodo Supplementary Material' referenced in this work. This document includes: all one-component gas-phase, aerosol, and surface retrievals for TOI-1685b (Supplementary Figures 1--3) and 55 Cancri e (Supplementary Figure 4--6); a best-fit components retrieval plot with all posteriors (Supplementary Figure 7); a surface geology plot using a uniform prior instead of the CLR prior (Supplementary Figure 8). The repository itself includes all surface albedos in the database, all the code used to produce the results and figures of this work, and the source code for \texttt{POSEIDON (v1.4)}. 

Online we provide open-source tutorial \texttt{Jupyter} notebooks \nt{titled `Rocky Planets with Reflecting and Emitting Surfaces', `Simulating Habitable Worlds Observatory Reflection Spectra', `Thermal Emission Retrievals of Bare Rocky Exoplanets', and `Advanced: Shiny Gray Decks in Eclipse Geometry'}, which are publicly available on the \href{https://poseidon-retrievals.readthedocs.io/en/latest/index.html}{\texttt{POSEIDON} online documentation}. \nt{Plots of albedo vs. wavelength for all entries in the surface albedo database can be seen in `Surface Albedo Database' page in page in \texttt{POSEIDON}'s \href{https://poseidon-retrievals.readthedocs.io/en/latest/content/opacity_database.html}{Opacity Database}.} 

\clearpage
\appendix

Figure \ref{fig:HWO} displays a proof-of-concept Habitable Worlds Observatory (HWO) retrieval discussed in \S \ref{sec:2.2.2}, Figure \ref{fig:turning-knobs} displays how the surface pressure was constrained in the retrieval results of Figure \ref{fig:retrieval-psurf}, \nt{and Figure \ref{fig:Linear-JWST-sims} shows retrievals results when using a uniform prior and linear percentages instad of a CLR prior and log percentages (compare to Figure \ref{fig:CLR-JWST-sims})}. \S \ref{Appendix-1} provides an in-depth discussion of the type of lab-measured reflectances included in the \texttt{POSEIDON} surface albedo database, as well as step-by-step derivations of equations used for airless rocky planet observed emission and reflection spectra. \S \ref{Appendix-2} contains supplementary geology tables and catalogs the surface albedo database. Table \ref{table:Geology Glossary} lists geology definitions and their exoplanet contexts, Table \ref{table:Surface Albedos Categories} documents the definitions of geology categories used to sort minerals and rocks in the albedo database and their Solar System context, and Table \ref{table:Surface Albedos Categories 2} further lumps the geology categories of Table \ref{table:Surface Albedos Categories} into detectable surface types, and potential considerations for how to interpret them if detected as an exoplanet surface. Table \ref{table:Surface Albedos} documents the primary surface albedos (directional-hemispherical reflectances) included in \texttt{POSEIDON} that cover wavelengths pertinent to reflection and emission \citep{Hu2012, Paragas2025, Hammond2025}, Table \ref{table:Basalt-Library} documents the basalt library of \citet{First2025}, Table \ref{table:Lava-Library} documents the lava-world surface library of \citet{Fortin2022}, and Table \ref{table:Surface Albedos Reflections} documents the surface albedos of \citet{GoodisGordin2024} that are relevant for reflection. See Table 1 in the Zenodo Supplementary Material (see \S \ref{sec:zenodo}) for documentation on the specific lab measurements of surface albedo entries (e.g., the \texttt{RELAB} reference number). 

\renewcommand{\thefigure}{A\arabic{figure}}
\renewcommand{\theHfigure}{A\arabic{figure}}
\setcounter{figure}{0} 
\setcounter{table}{0} 
\renewcommand{\thetable}{A\arabic{table}}
\renewcommand{\theHtable}{A\arabic{table}}

\newpage

\begin{figure*}[ht!]
    % \vspace{-1.3cm}
    \includegraphics[width=1.0\textwidth]{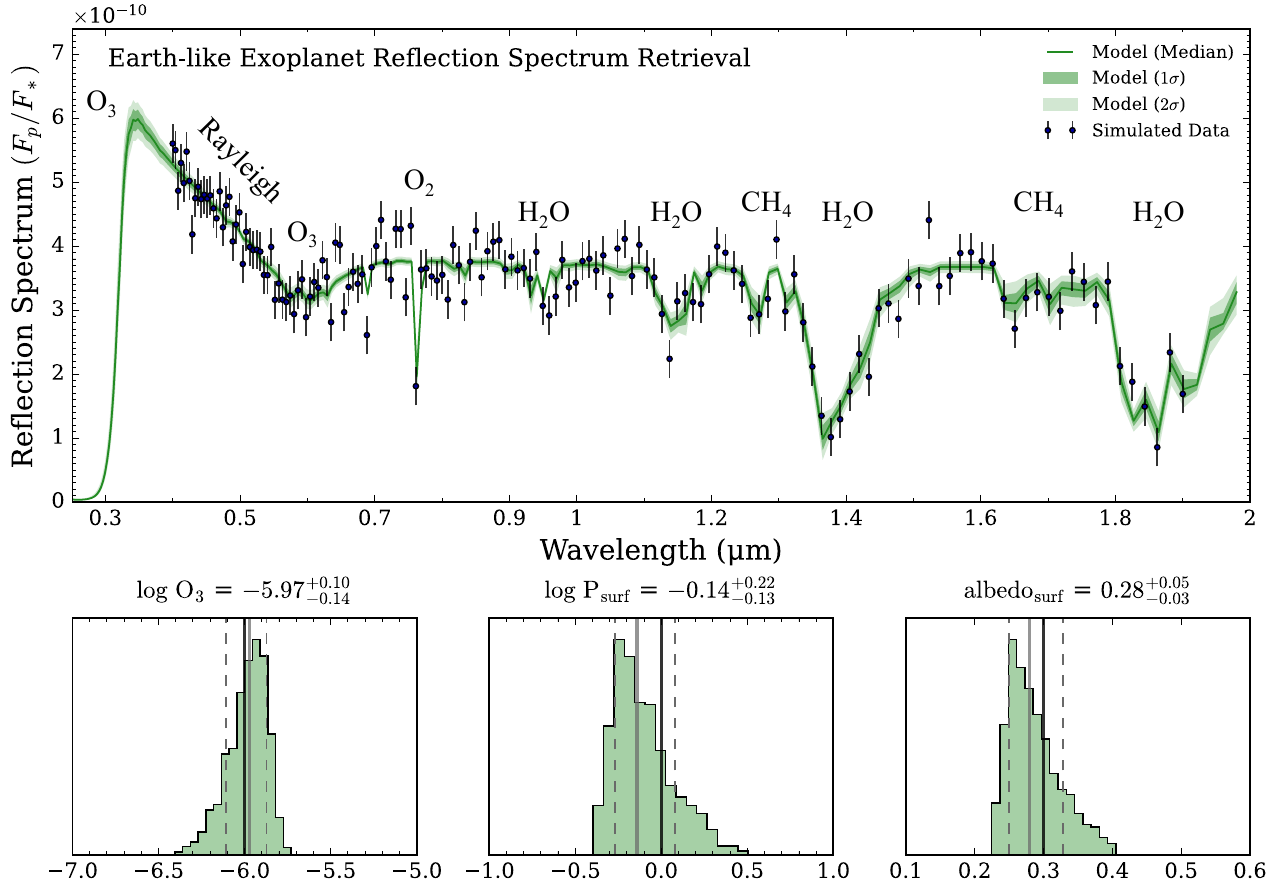}
    \caption{Proof-of-concept retrieval run on synthetic HWO gaussian-scattered data (SNR $\approx$ 10 at R = 100, error = 3e-11 ppm that is constant with wavelength), generated from a \texttt{POSEIDON} forward model assuming an Earth-like atmosphere (N$_2$ atmosphere with log O$_2$ = -0.68, log O$_3$ = -6, log H$_2$O = -4, log CH$_4$ = -5), and constant surface albedo (log P$_\mathrm{surf}$ = 0, albedo$_\mathrm{surf}$ = 0.3 ) with only reflected light (no thermal component). \texttt{POSEIDON}'s reflected light retrieval capabilities will be well suited for future HWO forward model and retrieval explorations. See \citet{Zelakiewicz2026} for a more in-depth application of \texttt{POSEIDON} on simulated HWO data.}     
    \label{fig:HWO}
\end{figure*}

\begin{figure*}[ht!]
    % \vspace{-1.3cm}
    \includegraphics[width=1.0\textwidth]{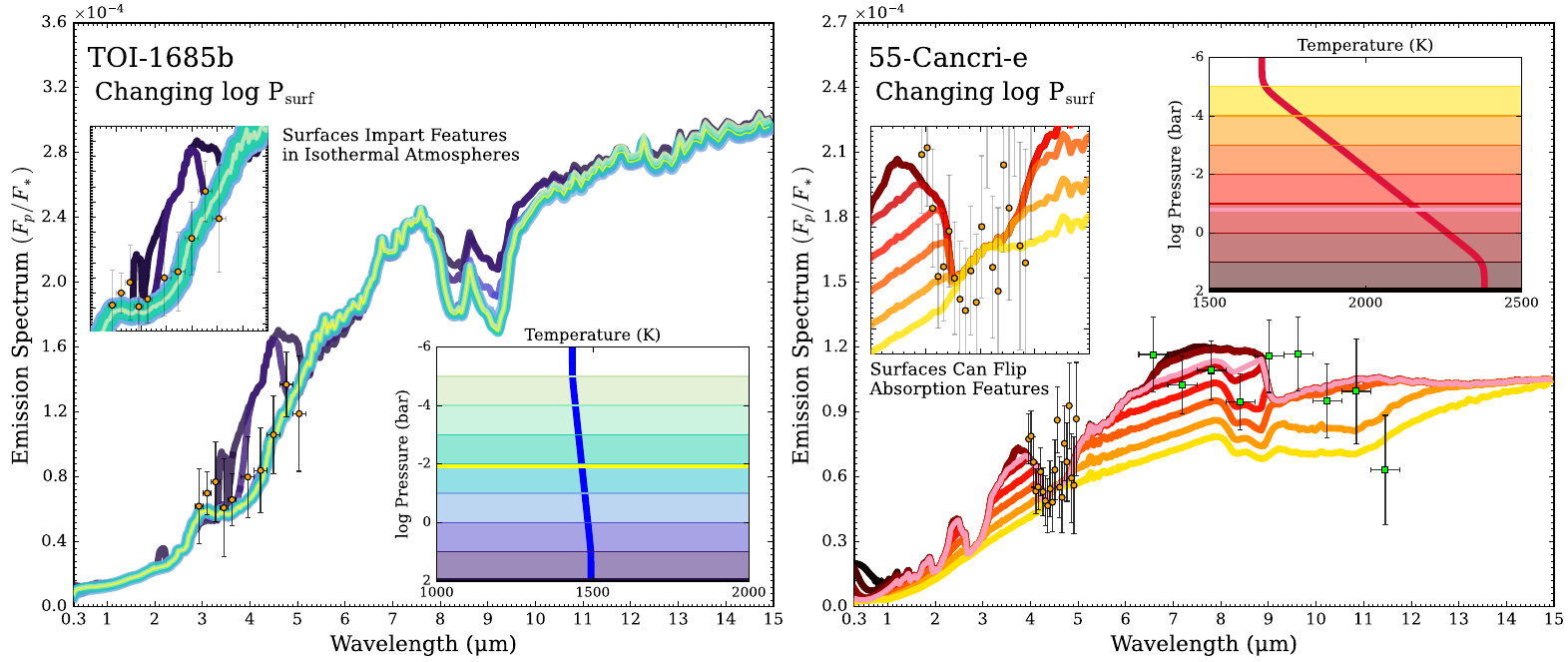}
    \caption{Display of how including surfaces with wavelength-independent albedos directly influences the resultant retrieved surface pressure (with ties to the specific features highlighted in Figure \ref{fig:main-takeaways}). We demonstrate this by taking the median retrieved parameters of the retrievals shown in Figure \ref{fig:retrieval-psurf} and varying only the surface pressure parameter. \textbf{Left} Changing the surface pressure for TOI-1685b, which retrieved a near-isothermal pressure-temperature profile (median retrieved spectra plotted in yellow, each surface pressure spectra corresponds to the color of horizontal lines in the pressure-temperature profile inset). As the surface pressure grows ($>$ 1 bar, corresponding to a thicker atmosphere), an N$_2$-N$_2$ pseudo-feature occurs in the 3-5 \textmu m region that is not favored by the data (see bottom left panel of Figure \ref{fig:main-takeaways}).  \textbf{Right} Changing the surface pressure for 55 Cancri e, which retrieved a gradient pressure-temperature profile (median retrieved spectra plotted in pink, each surface pressure spectra corresponds to the color of horizontal lines in the pressure-temperature profile inset). Low surface pressures ($\sim$ 1e-5 bars, corresponding to a thinner atmosphere) results in a CO$_2$ emission feature at 4.5 \textmu m due to a surface-atmosphere interface pseudo-temperature inversion that is not favored by the data (see bottom right panel of Figure \ref{fig:main-takeaways}). Additionally, the strength of the CO$_2$ absorption features (4.5 \textmu m, 9-15 \textmu m) and surface absorption feature (8-9 \textmu m) is controlled by the surface pressure, leading to a deep CO$_2$ absorption band and shallow surface absorption due to a \nt{thicker} atmosphere, that best fits the data (see top right panel of Figure \ref{fig:main-takeaways}).}    
    \label{fig:turning-knobs}
\end{figure*}

\begin{figure*}[ht!]
    % \vspace{-1.3cm}
    \includegraphics[width=1.0\textwidth]{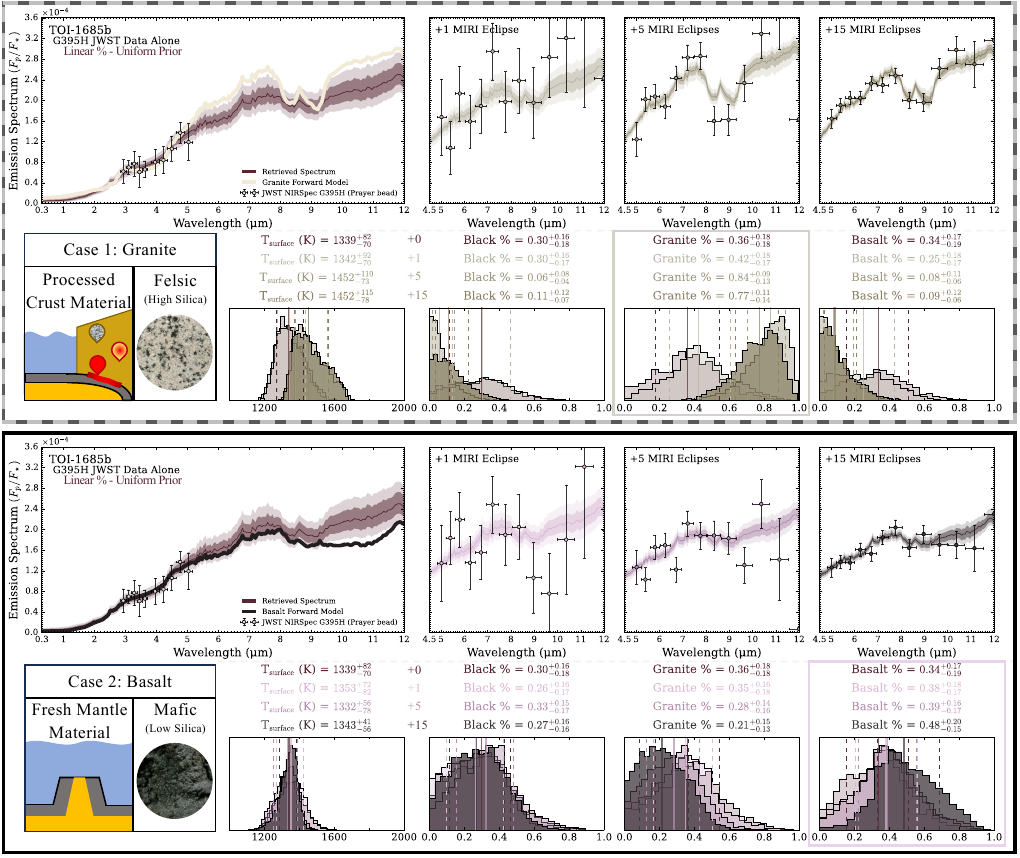}
    \caption{\nt{Same as Figure \ref{fig:CLR-JWST-sims}, but represents retreivals utilizing a uniform prior with linear surface component percentages in lieu of a CLR prior with log surface percentages. While the CLR prior excels in statistically detecting which surfaces are likely contributing the most to the resultant spectrum (`bulk' surface), a uniform prior with linear surface component percentages excels in capturing degeneracies that occur from linearly combining spectra from different surface albedo datasets. \textbf{Top} In 5 eclipses, granite is detected over basalt due to the feature at 8-9 \textmu m. \textbf{Bottom} While basalt is preferred at higher percentages than granite, due to a lack of SNR at 10-11 \textmu m and shape of the NIRSPec G395H data, granite is not entirely ruled out when using this prior. }}
    \label{fig:Linear-JWST-sims}
\end{figure*}

\clearpage

\section{Surface Albedos and Modeled Bare Rocky Surface Spectra}\label{Appendix-1}
In this section, we define the lab measured surface albedos in our database as well as a step-by-step derivation of the observed, bare rocky surface emergent thermal emission and reflected flux spectra. 

% \citet{Paragas2025} and \citet{First2025} directly measure the directional-hemispherical reflectances ($r_\mathrm{h}$), the open-source atmospheric retrieval code \texttt{Platon v6.2}\footnote{https://github.com/ideasrule/platon}\citep{Zhang2019, Zhang2025, Paragas2025} reports the hemispherical reflectances and geometric albedos of the surfaces from \citet{Hu2012}. \citet{Hammond2025} reports the single scattering albedo of their compiled surfaces\footnote{https://zenodo.org/records/14017134} with equations to convert to directional hemospheric albedo, spherical albedo, and geometric albedo. 

In \texttt{POSEIDON}, all surfaces in the database are in their directional-hemispherical reflectance form (with the exception of the albedos featured in Table \ref{table:Surface Albedos Reflections}). The directional-hemispherical reflectance (sometimes notated as `black (clear) sky albedo', \citealt{Wang2004}, and often used interchangeably with `hemispherical reflectance', r$_\mathrm{dh}$) is the total fraction of light scattered in all emergent directions (in the upward direction) by a surface under direct illumination from above by a highly collimated source \citep{Hapke2012}. As defined in \citet{Hapke2012}, the first signifier `directional' refers to the collimation of the source while the second signifier `hemispherical' refers to the collimation of the measurement. Directional-hemispheric reflectance is the most commonly reported lab-measured reflectance since most commercial instruments measure it. Its use in planetary science is ubiquitous since incident sunlight (the `source') can often be thought of as highly collimated (`directional'), while integral derivations of flux often require the total amount of flux reflected in all upwards directions (`hemispherical'). 

We can compare this to bi-hemispherical (i.e., hemispherical-hemispherical) reflectance, or `white (cloudy) sky albedo', which is the total fraction of light scattered upwards from diffuse illumination (diffuse being light scattered to a surface from atmospheric gases or aerosols, generally assumed to be isotropic). In general, white-sky diffuse albedo is slightly larger than black-sky direct albedo due to the diffuse illumination being near-isotropic \citep{Oleson2003}, allowing secondary surface effects such as angle-dependent reflection, multiple scattering, forward scattering, surface geometry, and others to allow more light to be reflected back to the detector. 

Some sources also report bi-directional (i.e., directional-directional) reflectance which is dependent on the angle of the detector (and is analogous to the reflected light from the surface our eyes perceive on a clear, sunny day). We urge users of \texttt{POSEIDON} caution in using bi-directional reflectances, since to accurately derive the observed flux, multiple emergent angles would need to be measured and included in the integrals below \citep{Gkouvelis2025two}. However, if a medium is generally considered specular (i.e., doesn't scatter much reflected light, like a perfect mirror), the directional-hemispherical reflectance is roughly equivalent to the directional-directional reflectance. Many databases with reflectance measurements made for Earth and Solar System studies, where the incident and observing angle is exactly known and objects aren’t always observed at opposition, report bi-directional reflectances. When the reflectance factor (denoted REFF in \citealt{Hapke2012} and \citealt{Hammond2025}) is provided, one can convert from bi-directional reflectance to directional-hemispherical reflectance;  an example of this are the \texttt{RELAB Spectral Database}\footnote{https://sites.brown.edu/relab/relab-spectral-database/} \citep{Milliken2021} measurements, which are typically a collection of bi-directional visible reflectances and bi-conical near- and mid-infrared reflectances. These measurements can be converted to bi-directional reflectances, then to single scattering albedos, and finally directional-hemispherical reflectances following the procedure of \citet{Hapke2012} and the \texttt{RELAB} manual, which is explicitly utilized in \citet{Hammond2025}. In the case where only bidirectional lab data is available for a specific use-case (with no reported REFF), we urge users to make explicit note of this in their publications and that their surface albedos might be under-estimating reflectance.

The single scattering albedo ($\omega$), which is reported in \citet{Hammond2025}, describes the probability of a single particle's interaction with light (probability of scattering or absorbing a single incident photon). The single scattering albedo can be converted to to directional hemispherical reflectance via \citep{Hapke2012}

\begin{align}
r_h(\lambda) &= \frac{1-\gamma(\lambda)}{1 + 2\gamma(\lambda)\mu_0}\\
\gamma(\lambda) &= \sqrt{1-\omega(\lambda)}
\end{align}

\noindent where $\mu_0$ = 1 assuming solar zenith.

For planetary objects, we are often interested in disc-averaged albedo definitions. Here we repeat the definitions of bond albedo, spherical albedo,  and geometric albedo of \citet{Heng2021}: the bond albedo is the total fraction of starlight reflected by a planetary object averaged across all viewing angles and wavelengths, the spherical albedo is the wavelength-dependent albedo averaged across all viewing angles, while the geometric albedo (A$_g$) is the wavelength-dependent albedo measured when an exoplanet is at opposition (star is between observer and exoplanet, orbital phase angle of 0). For our following derivation of observed exoplanet flux where we assume we are observing at secondary-eclipse (see \citealt{Seager2010, Heng2021} for phase-resolved solutions), we need to relate lab-measured directional-hemispherical reflectance to planetary geometric albedo. 

In order to model the disc-averaged spectra of exoplanets measured by observatories, we utilize lab-measured directional hemispherical reflectances and assume we are observing the planet at occulation (secondary-eclipse). We then compute the bare-rock thermal emission and reflection by assuming most of the observed emission (in W m$^2$ m$^{-1}$) comes from the planet's substellar point (i.e., we take into account the planet's spherical shape when computing the observed flux, \nt{see \citet{Hu2012}}):

\begin{equation}
    F_\mathrm{bare\;rock,obs} = \left(\frac{R_\mathrm{p}}{d}\right)^2 \int_{-\frac{\pi}{2}}^{+\frac{\pi}{2}} \int_{-\frac{\pi}{2}}^{+\frac{\pi}{2}} I_\mathrm{p}(\theta,\phi)\mathrm{cos}^2(\theta)\mathrm{cos}(\phi)d\theta d\phi
\end{equation}

\noindent where $R_\mathrm{p}$ is the radius of the planet, $d$ is the distance to the system, and the latitude-longitude dependent intensity $I_\mathrm{p}(\theta,\phi)$ (with the substellar point at $\theta$ = 0 and $\phi$ = 0) is composed of a reflected ($I_\mathrm{reflected}(\theta,\phi)$) and a thermal emission ($I_\mathrm{thermal}(\theta,\phi)$) component

\begin{equation}
    I_\mathrm{reflected}(\theta,\phi) = r_\mathrm{dh}(\mu_0,\lambda)F_{\mathrm{s}}(\theta,\phi,\lambda) \left(\frac{R_\mathrm{s}}{a_\mathrm{p}}\right)^2\frac{\mu_0}{\pi}
\end{equation}

\begin{equation}
    I_\mathrm{thermal}(\theta,\phi) = \epsilon_d(\mu,\lambda)B(T(\theta,\phi),\lambda) \; .
\end{equation}

\noindent Where $\mu_0$ is the cosine of the incident angle (denoted as cos(0$^\circ$) = 1 for light incident perpendicular to the surface), $\mu$ is the emergent angle, and $F_{\mathrm{s}}(\theta,\phi,\lambda) (\frac{R_\mathrm{s}}{a_\mathrm{p}})^2$ is the incident stellar flux at the planet's orbital distance from the star. The directional emissivity ($\epsilon_\mathrm{d}$) is given by Kirchoff's Law ($\epsilon_\mathrm{d}$ = 1 - $r_{\mathrm{dh}}$). By utilizing the directional-hemispheric reflectance/directional-emissivity measurement, we lose the $\mu$ dependence. Additionally, we assume a uniform surface temperature ($T_\mathrm{surf}$) across the entire planet. Therefore, the observed emitted thermal flux becomes

\begin{align}
    F_\mathrm{bare\;rock,thermal,obs} &= \left(\frac{R_\mathrm{p}}{d}\right)^2 \int_{-\frac{\pi}{2}}^{+\frac{\pi}{2}} \int_{-\frac{\pi}{2}}^{+\frac{\pi}{2}} \epsilon_d(\lambda)B(T_\mathrm{surf},\lambda) (\theta,\phi)\mathrm{cos}^2(\theta)\mathrm{cos}(\phi)d\theta d\phi \\
    &= \left(\frac{R_\mathrm{p}}{d}\right)^2 \; \pi \; \epsilon_d(\lambda)\;B(T_\mathrm{surf},\lambda) \; .
\end{align}

To convert to the observed $\frac{F_\mathrm{p}}{F_\mathrm{s}}$, we divide by $F_{\mathrm{s,obs}}$

\begin{equation}
F_{\mathrm{s,obs}} = \left(\frac{R_s}{d}\right)^2* F_s
\end{equation}

which results in 

\begin{equation}
\frac{F_\mathrm{bare\;rock,thermal,obs}}{F_{\mathrm{s,obs}}} = \left(\frac{R_\mathrm{p}}{R_\mathrm{s}}\right)^2 \; \frac{\pi \; \epsilon_d(\lambda)\;B(T_\mathrm{surf},\lambda)}{F_s} \; .
\end{equation}

Both \citet{Paragas2025} and \citet{First2025} follow this approach in their bare-rock spectral models. This approach is identical to the approach we take in this work with \texttt{POSEIDON}, for both our bare-rock and surface-atmosphere models. While our bare-rock approach is identical to the one above, in our surface-atmosphere model derived from the Toon scattering modules of \texttt{PICASO} we still use directional-hemispherical reflectances/directional emissivities as surface albedo inputs since the emergent and incident angles are applied in the computations of those functions alongside the Gauss angle approximation. We also assume direct instead of diffuse radiation, even though the models can include both clear and cloudy atmospheres. Future work could investigate `blue-sky albedo \citep{Wang2004}' which combines both direct (directional-hemispherical) and diffuse (bihemispherical) reflectivities. 

Assuming that $r_\mathrm{dh}$ doesn't depend on incident angle, and F$_s$ doesn't depend on latitude and longitude, 

\begin{align}
    F_\mathrm{bare\;rock,reflected,obs} &=\left(\frac{R_\mathrm{p}}{d}\right)^2 \int_{-\frac{\pi}{2}}^{+\frac{\pi}{2}} \int_{-\frac{\pi}{2}}^{+\frac{\pi}{2}} r_\mathrm{dh}(\lambda)F_s(\lambda)\left(\frac{R_\mathrm{s}}{a_\mathrm{p}}\right)^2\frac{\mu_0}{\pi}\mathrm{cos}^2(\theta)\mathrm{cos}(\phi)d\theta d\phi \\
    %&= (\frac{R_\mathrm{p}}{d})^2 r_h(\lambda)F_s(\lambda)(\frac{R_\mathrm{s}}{a_\mathrm{p}})^2 \\
    &= \left(\frac{R_\mathrm{p}}{d}\right)^2 r_\mathrm{dh}(\lambda)F_s(\lambda)\left(\frac{R_\mathrm{s}}{a_\mathrm{p}}\right)^2 \frac{1}{\pi} \int_{-\frac{\pi}{2}}^{+\frac{\pi}{2}} \int_{-\frac{\pi}{2}}^{+\frac{\pi}{2}}\mathrm{cos}^3(\theta)\mathrm{cos}^2(\phi)d\theta d\phi \\
    &= \left(\frac{R_\mathrm{p}}{d}\right)^2 r_\mathrm{dh}(\lambda)F_s(\lambda)\left(\frac{R_\mathrm{s}}{a_\mathrm{p}}\right)^2 \frac{1}{\pi} \frac{2\pi}{3} \\
    &= \left(\frac{R_\mathrm{p}}{d}\right)^2 r_\mathrm{dh}(\lambda)F_s(\lambda)\left(\frac{R_\mathrm{s}}{a_\mathrm{p}}\right)^2 \frac{2}{3} \\
\end{align}

\noindent where $\mu_0 = \mathrm{cos}(\theta)\mathrm{cos}(\phi)$ at opposition \citep{Hu2012}. Dividing by $F_{\mathrm{s,obs}}$ results in 

\begin{align}
    \frac{F_\mathrm{bare\;rock,reflected,obs}}{F_{\mathrm{s,obs}}}&= \left(\frac{R_\mathrm{p}}{a_p}\right)^2 r_\mathrm{dh}(\lambda) \frac{2}{3} \\
    &=\left(\frac{R_\mathrm{p}}{a_p}\right)^2 A_\mathrm{g}(\lambda)
\end{align}

\noindent where $A_\mathrm{g}(\lambda)$ is the geometric albedo.

%\noindent where $\mu_0 = cos(i) = cos(\theta)cos(\phi)$ \citep{Hu2012} and we assume that $\mu_0 = 1$ (corresponding to flux incident perpendicular to the surface). If the $\theta$ and $\phi$ dependence is included, the expression above gains an additional factor of $\frac{2}{3}$ (corresponding to the geometric albedo of a Lambertian sphere). Following \citet{Seager2010}, the spherical albedo is defined as the fraction of incident stellar radiation scattered back into space in all directions (the quantity we are interested in) is equal to $\frac{3}{2}$ of the geometric albedo, which gets rid of the factor of $\frac{2}{3}$.

We note that our reflected flux approach differs from the equations of \citet{Paragas2025} by a factor of $\frac{2}{3}$. \nt{The factor of $\frac{2}{3}$ comes from the relationship between the Bond albedo and geometric albedo for Lambertian surfaces, where for a Lambert surface the value is $\frac{3}{2}$. While this assumption is essential for the derivation, we note that all solar system planets have a ratio less than $\frac{3}{2}$ (e.g., 0.6 for Mercury, 1.1 for Venus, 0.7 for Earth, 0.9 for the Moon, and 0.34-0.54 for asteroids, \citealt{Seager2010}).} Our models also do not include any secondary effects, such as thermal beaming, opposition surge enhancement that can occur at small phase angles \citep{Gkouvelis2025one}, surface morphology (`roughness') and crater shadows, or latitude-longitude dependent surface temperatures (though, it has been shown that opposition surge is a minor effect that requires high SNR time-resolved spectra to resolve, \citealt{Jones2025}). Future work should investigate these secondary effects if they are shown to affect JWST and HWO observations (see \S \ref{sec:6.1}).

\clearpage

\section{Supplementary Geology Tables and the \texttt{POSEIDON} Surface Albedo Database}\label{Appendix-2}

\setlength{\extrarowheight}{1.2pt}
\startlongtable
\begin{deluxetable*}{lp{8cm}p{5.5cm}}
\colnumbers
\tablecolumns{3}
\tablewidth{0pt}
\tabletypesize{\footnotesize}
\tablecaption{Supplementary Geology Glossary\label{table:Geology Glossary}}
\tablehead{Word & Definition & Exoplanet Context}
\startdata
Mineral & Inorganic element or compound with a orderly internal structure and characteristic chemical composition. See succint introduction on minerals, \citet{Vaughan2014}. & An exoplanet's observed disc-integrated dayside being composed of pure minerals is unlikely, though the surface can have rocks with large mineral inclusions (e.g., phenocrysts). Depending on the rock, distinct spectral signatures of specific minerals (like quartz in granite) can be readily detected. \\
Rock & An aggregate of one or more minerals. Rocks are determined by their mineralogy and bulk chemical composition, where resultant spectra is determined by mineralogy (i.e., two rocks of the same chemical composition can have different spectra due to different mineralogy, \citealt{Paragas2025}). See succinct introduction on rocks, \citet{Zalasiewicz2016} (Z16). & If the observed dayside of a rocky planet is completely cooled and solid, we are likely to observe spectral signatures of specific rocks (or, a conglomeration of rocks). The resulting color and spectral features of a rock is largely given by its mineral composition, though typically minerals do not spectroscopically mix in a linear fashion \citep{Hu2012,Ehlmann2012}. \\
\hline
Igneous Rocks& Rocks formed from the cooling of lava (extrusive/volcanic) or magma (intrusive/plutonic). They are the most common rock found on the surface of rocky planets in the Solar System, and are the first to form as a rocky planet cools from its initial formation state (magma ocean) and from resurfacing a planet's surface from subsequent partial melts of mantle material.$^{\mathrm{Z16}}$  & With the exception of Earth, igenous rocks are the most common surface rock for Solar System rocky planets. The surfaces of currently observable rocky exoplanet daysides will be hot enough that we expect them to be molten or cooled molten, they therefore compose a majority of our database and the variety of processes that form them are detailed below in this table.\\
Sedimentary Rocks& Rocks formed from accumulation of pre-existing rocks over time. They are the most common rocks on Earth's surface, due to its unique processes like plate tectonics (that pushes portions of the crust high enough into the atmosphere to erode), atmospheric processes (winds), surface water (oceans and rivers), and glacial processes. Some common sediments that form from igenous rocks include clays, muds, sand (mostly composed of quartz), and gravel; salts and calcium carbonates are also considered sedimentary and precipitate directly out of ocean water. Common rocks that form from the lithication of sediments are mudrocks (e.g., shale), sandstone, and limestone. Surface water and limestones have an inverse relationship with atmospheric carbon dioxide in the Solar System: Earth's carbon dioxide is stored in limestone while Venus, which has no limestone, has a carbon dioxide dominated atmosphere \citep{Hunten1993}. Sedimentary rocks can indicate the presence of past (sedimentary rocks have been found in proposed river basins on Mars \citep{Lewis2014}) or present liquid water, or just any process (like atmospheric erosion) that can create and accumulate sediment over time (i.e., dunes formed by winds on Mars, or see potential Venus sedimentary processes,  \citealt{Carter2023}).$^{\mathrm{Z16}}$ & Sedimentary rocks require atmospheres and/or liquid water to form. Sedimentary rocks can form in hot and dense atmospheric environments (similar to Venus) that could be present in currently observable exoplanets. They could also possibly be present on airless rocky exoplanets that once housed a substantial atmospheres and/or surface water. The database presented in this work includes a few clays and salts.\\
Metamorphic Rocks& Rocks formed from transformation of other rocks due to high heat and/or pressure, or hydrothermally. In Earth's traditional rock cycle, sedimentary rocks formed from igenous rocks get buried over time (i.e., in mountain belts) where, due to higher pressures and temperatures (regional metamorphism) mica minerals form from mudstones (that then can metamorphize a variety of rocks, like slate, schist, and gneiss, where gneiss is nearly identical in composition to the igenous rock granite), sandstone forms quartzite, and limestone forms marble, before finally melting into a magma to form new igenous rocks. High-pressure, low-temperature metamorphism that occurs at subduction zones in the ocean (hydrothermal) can form the mineral serpentine. Contact metamorphism occurs when rock is heated and crystallized due to heating from nearby magma. Impact metamorphism is where large impacts form glasses and high-pressure minerals. Metamorphic rocks are present on the surface of Earth due geologic uplift (via plate tectonics) and subsequent erosion of overlying layers.$^{\mathrm{Z16}}$ Metamorphic rocks could be the natural form of rocks of a surface with an overlaying thick and high temperature atmosphere, as in the case of Venus (though, chemical alteration via sulfuric acid might dominate Venus's metamorphic processes) \citep{Barsukov1982,Semprich2025}. Hydrothermal and impact metamorphic rocks have been found on Mars \citep{Hazen2023}. & For rocky, airless exoplanets with magma oceans at their substellar point, the solidified rock around the magma ocean might be primarily in the form of metamorphic rocks from contact metamorphism with the neighboring magma ocean. On close-in rocky exoplanets, a form of high-temperature, low-pressure metamorphism might occur on the dayside due to incident irradiation \citep{Chao2021}. Exoplanets with Venusian like atmospheres might have chemical alteration of the surface. Evidence of past bombardment can form specific rocks via impact metamorphism. Past surface water can also be probed via the presence of hydrothermal metamorphic minerals (like serpentine). While these are an incredibly important class of rock, the database presented in this work does not include any explicit metamorphic rocks, but does include some rocks with serpentine. \\
\hline
Intrusive vs Extrusive & Many igneous rocks have different names for intrusive vs extrusive variants of rocks with identical chemical compositions (and therefore, similar absorption features in their spectra). The main difference between the two variants are their textures, where intrusive rocks are typically coarse grained due to slow-cooling (which usually occurs underground, which allows large crystals to grow) while extrusive rocks are typically fine-grained due to rapid cooling (usually due to lava interfacing with an atmosphere or surface water, causing it to rapidly cool).$^{\mathrm{Z16}}$ & The difference in texture can result in differences in observed albedos and spectra \citep{Paragas2025,Hammond2025}. Where relevant, we will indicate when two geological categories in our database are an intrusive/extrusive pair (examples being: granite/rhyolite, diorite/andesite, gabbro/basalt), though we note that care must be taken when applying Earth-like definitions of intrusive vs extrusive to exoplanet applications.  \\
\hline
TAS Diagram & The Total Alkali-Silica (TAS) diagram categorizes extrusive igenous rocks based on alkali metal oxide (Na$_2$O+K$_2$O) and SiO$_2$ content to distinguish whether they formed from alkaline-rich/poor or silica-rich/poor magma. For a textbook review of magma petrology, see \citet{Winter2014} (W14) or \citet{Philpotts2022}. & While not directly applicable to exoplanets, many rocks are explicitly classified on this diagram. When relevant, we will indicate when a geological category in our database is a specific rock category on the TAS Diagram. \\
Alkaline Content in Magma & On Earth, alkali-rich magma is thought to form deep in the mantle, and found at areas of continental rifts and hotspots, while alkali-poor magma is thought to form from shallow melting of the mantle and is found at mid-ocean ridges where it cools to form the most common igneous rock on Earth (sub-categorized as tholeiitic (iron enriched) and calc-alkaline (no iron enrichment)).$^{\mathrm{W14}}$ & Alkali oxides, by themselves, do not have strong mid-infrared features, and therefore are not readily probed with JWST. However, feldspars (which contain alkalis + silica) which compose many igenous rocks can be probed by their mid-infared features. \nt{See \citet{vanBuchem2026} to see how changes in alkali content in lava can change atmospheric emission spectra.} \\ 
Silica Content in Magma & The silica content of magma is mostly determined by the age of magma and its degree of cooling (as magma cools, dense crystallized refractory material sinks and the remaining magma is enriched with silica). It is also used to determine whether or not a magma is mantle-derived (basaltic) or a crustal (granitic) melt. The silica content of magma strongly determines the viscosity (higher silica content results in a higher viscosity/more thick magma), and therefore determines how volcanoes erupt.$^{\mathrm{W14}}$ Silica content can be used to further categorize rocks into ultramafic, mafic, intermediate, and felsic categories (see Table \ref{table:Surface Albedos Categories 2}). & Silica content is the most accessible characteristic of rocky exoplanet surfaces that can derived from spectroscopy with JWST due to strong mid-inrared Si-O absorption and the Christiansen Feature. In our database, we choose to focus on categorization via silica content for this reason. Since silica content gives a wealth of degenerate information about the surface, such as initial formation abundances and degree of partial melting, care must be taken in interpreting it.  \\
\hline
Mantle Material & The mantle for evolved planets (with a fully cooled crust, and a fractionated core) is the middle region of a terrestrial body (in contrast, for lava worlds without a cooled crustal region the observable surface is often referred to as the mantle). Surface geology is intrinsically linked to a planet's mantle chemistry, wherein partial melts of mantle material cools and is reprocessed over time to form the surface of a planet. Typical mantle mineraology and composition can be modeled from stellar compositions (see \S \ref{sec:5.1}). On Earth, 90\% of the upper mantle is composed of olivine, orthopyroxene, and clinopyroxene minerals that form mostly peridotite (and some pyroxenite) rocks.$^{\mathrm{Z16}}$ & Earth-like upper mantle rocks (composed of olivines and pyroxenes) are predicted to be the dominant mantle material for exoplanets from models using the \texttt{Hypatia Catalog} \citep{Putirka2019,Wang2022}. However, more exotic mantle chemistry has been observed from white-dwarf pollution measurements where quartz and periclase saturation can instead occur \citep{Putrika2021}. While the detection of surface composition can hint at a parent mantle composition, white dwarf pollution measurements are the only method by which to directly measure the mantle compositions of exoplanets \citep{Xu2021}. \\
Partial Melts  & As a terrestrial body cools from its initially molten state, refractory material with high melting points (i.e., olivines, pyroxenes) begin to solidify and remove themselves from the melt mixture, changing the bulk composition of the remaining melt. If the solidified material is removed from the system (i.e. sequestered deeper in the body via differentiation), the resulting melt will cool into a rock with a different composition from the original melt. The reverse process is also true, where as solidified mantle rocks are heated, volatile material begins to melt. Both processes above play a large role in determining the surface composition$^{\mathrm{Z16}}$ & Partial melts of solid mantle material in Earth (i.e., ultramafic peridotites) produce more silica-rich magmas that cool to form mafic basalts. This pattern extends to all Solar System rocky planets, and is predicted to extend to most exoplanets. In general, more silica-rich material is more volatile and is first to melt (last to solidify), which is important to take into account when predicting the surface composition of planets with more exotic mantle compositions.  \\
Melt fraction & The percentage of a rock that has melted from an original rock composition, where the resulting partial melt is typically more Si-rich than the original rock composition. & For lava worlds, the composition of the observable lava on the dayside will have a composition that is a melt fraction of the mantle. The quenched glasses from \citet{Fortin2022} represent the resulting melt chemistry of a mantle that has been melted a certain percentage and then rapidly quenched (to retain the internal structure of the magma). Magma composition, like solidified rocks, can be probed spectroscopically. \\
%Tholeiitic Magma & Magnesium-rich magma.\\
Oxygen Fugacity & $fO_2$ quantifies the amount of available oxygen to take part in oxidizing (loss of electrons/gaining oxygen) or reducing (gain of electrons/losing oxygen) chemical reactions in the interior of a planet. For internal geology, $fO_\mathrm{2}$ is proportional to Fe$^{\mathrm{+3}}$ (ferric) / Fe$^{\mathrm{+2}}$ (ferrous) abundances. Low $fO_2$ favors metallic iron (Fe) and w\"ustite (FeO), while high fugacity favors hematite (Fe$_\mathrm{2}$O$_\mathrm{3}$) \citep{Cortell2025}. Oxygen fugacity controls the amount of interior metallic iron (and therefore the size of the iron core) \citep{Unterborn2017}, the ease by which rock melts to produce magma (with higher fugacities resulting in a lower melting point) \citep{Lin2021b}, and which molecular species are out-gassed \citep{Gaillard2021} (i.e., H$_\mathrm{2}$ vs. H$_\mathrm{2}$O, CO vs CO$_\mathrm{2}$, H$_\mathrm{2}$S vs SO$_\mathrm{2}$). In the Solar System, Earth's mantle is the most oxixided while Mars and the Moon are more reduced (where, \citet{Deng2020} argue that deeper magma oceans on larger planets naturally result in a more oxidized surface). While the above concerns mantle oxygen fugacity, the atmosphere of a planet can also reduce or oxidize crustal material (see Oxidized vs Reduced Surface in Table \ref{table:Surface Albedos Categories 2}).
& While the oxygen fugacity of an exoplanet's mantle cannot be directly known, many studies have explored how detecting specific atmospheric compositions can be tied to a mantle's oxygen fugacity \citep[e.g.,][]{Drant2025,Perez2025,vanBuchem2026}. White dwarf pollution studies have shown that many exoplanets probably form with Earth-like oxygen fugacities \citep{Doyle2019}. While the amount of reactive oxygen in the mantle is directly tied to its formation conditions, the actual mantle oxygen fugacity can change depending on a planet's mineralogy (which depends on the initial Mg/Si from formation), wherein suppressed formation of a mineral like pyroxene can inherently increase mantle oxygen fugacity \citep{Guimond2023} and change which molecular species are outgassed.  \\
\hline
Crustal Material  & For terrestrial planets with a solid and cooled surface, the crust represents the top-most layer of the planet. The resulting bulk composition of the crust is a strong function of time from a planet's initial formation and is a complex result of cooling processes (like differentiation), resurfacing processes (like plate tectonics, volcanism, and large impacts), and alteration processes (like aqueous alteration and atmospheric oxidation). Here, we include the the crustal definitions of \citet{Condie2022} (C22) that outline the types of crusts in the Solar System. & For exoplanets with cooled crusts, their dayside surface composition will be a strong probe of past and current geologic processes. \\
Primitive Crust & The crust that forms first from planetary accretion, which is common on smaller asteroids and comets, but rarely survives on larger bodies (due to their large heating during formation, which forms a magma ocean). An example of a primitive crust is on Vesta.$^{\mathrm{C22}}$ & This crust rarely survives on larger bodies, and therefore is not expected to be observed on exoplanets. \\
Primary Crust  & The crust that forms on a planet due to the differentiation and solidification of an initial magma ocean. While rare in the Solar System, there are examples of isolated regions on terrestrial bodies with primary crust: e.g., the lunar highlands, which formed from rocks (i.e., anorthosite) composed of plagioclase feldspars (a mineral group with the endmember minerals albite and anorthite) that crystallized in the magma ocean and floated to the top.$^{\mathrm{C22}}$ & For young exoplanets, or ones without active geology, their crust might be primary. As an initially molten surface cools, dense condensed material that crystallizes out is sequestered deeper in the mantle (olivines, pyroxenes) while less-dense condensed material (like anorthosite) floats to the top. A similar analysis done by modeling the resulting top-level crust from cooling a compositionally specific magma ocean can be done to determine whether or not an observed crust is primary or not. \\
Secondary Crust  &  Produced from partial melts of solid mantle material (which are typically ultramafic rocks) that usually cools into mafic basalts. This form of crust is the most common on rocky planets in the Solar System, and replaces the primary crust over time due to mantle overturn and volcanism. On Earth, fresh secondary crust is formed at mid-ocean ridges.$^{\mathrm{C22}}$ & Most exoplanets will probably have some form of secondary crust that has been formed from the cooled material produced via the partial melting of their solid mantle. Additional processes, like impacts, intense stellar irradiation, and past/present atmosphere might further alter the secondary crust. Exoplanets without active geology will have a more altered secondary crust. \\
Tertiary Crust  & Requires the gradual modification of the secondary crust through a process like plate tectonics. The only known example is the continental crust of Earth.$^{\mathrm{C22}}$ & Earth's thick and old tertiary crust is formed from plate tectonics, which are solely unique to Earth and require large amounts of surface water to occur (water lowers the melting point of mantle material, lowers viscosity, and reduces friction between plates,  \citealt{Tikoo2017}).$^{\mathrm{Z16}}$ While it is unknown whether or not observable exoplanets will have plate tectonics, it is possible that the idea of what constitutes a tertiary crust will expand with the as-of-yet unknown geologic processes that dominate in highly-irradiated, tidally locked rocky exoplanets. \\
\hline
Surface Composition & The resulting surface composition of a terrestrial body, and therefore what can be detected directly with spectroscopy, is determined by a complex interplay of a planet's initial formation composition, the cooling of the planet over time (which determines the level of differentiation and resurfacing processes that occur), the heating of the planet (which in the case of tidally-locked worlds with dayside temperatures above the melting point of the exoplanet's crustal rock composition, can result in dayside magma oceans), whether or not the planet is still geologically active (i.e., Mars due its small size is now geologically dead), and subsequent alteration of the surface from a past or present atmosphere, liquid water, or impacts. & While these processes are difficult to disentangle from the detection of geology on a exoplanet, inferences can be made about what lead to a specific observed surface composition (see \S \ref{sec:5.1} and Table \ref{table:Surface Albedos Categories 2}). For example, the detection of a granitic surface could indicate the presence of current or past Earth-like plate tectonics, the cooling of a planet that formed with a silica-rich mantle (an `exotic' mantle), or the resurfaced metamorphic transformation of rock into a mineralogical form similar to granite (e.g., gneiss). The database presented in this work focuses on lab data of rocks with specific mineralogies and chemical compositions, where if the spectral signature of one is detected, how that rock (or something similar to it) is the primary spectral signature of an exoplanet's surface can be deducted within the context of modern-day geology. Other works have explored the detection of an exoplanet's chemical composition in lieu of a specific rock (SiO$_2$ wt\% content via detection of the Christiansen Feature,  \citealt{Paragas2025,First2025}), and methods by which to learn more about the surface composition from observed atmospheric compositions \citep{Herbort2025}. \\
%\hline
%Formation  & When a terrestrial planet first forms it is purely molten rock and all material (e.g., Fe, Si, Mg) is uniformly mixed. This initial molten composition is determined by protoplanetary disc composition and distance from the host star. Previous work \citep[i.e.,][]{Putirka2019} has linked predicted mantle compositions with refractory host star abundance ratios. In non-differentiated bodies, the composition of the surface can be tied directly to its initial formation composition, where the rocks that end up forming are due to the internal and external temperature and pressure conditions unique to the planet.\\
%Differentiation  & As a planet cools, dense materials (like Fe) cool and sink, where this cooling rate depends on a planet's incident irradiation and the sinking of material is linked to the planet's gravity. Depending on the initial composition and gravity, it will form distinct material layers (i.e., core, mantle, crust; though not all differentiated bodies have iron cores). No matter the initial composition, the bulk surface of a differentiated body will represent the lightest material that was last to solidify (typically silica-rich material for known rocky bodies). \\
%Alteration  & Alteration is any process that can change a rock's composition after it has formed from cooled molten rock. This includes processes that involve an atmosphere (erosion, oxidation), interactions with liquids (aqueous alteration, erosion), and high-energy events like impacts. Processes like plate tectonics can also recycle rocks and reprocess them.\\
\hline 
Resurfacing & The resurfacing of a planet via melted mantle material, and therefore the formation of secondary crust (which produces basalts in all the Solar System rocky planets) can occur in many different ways; on Mercury, the youngest surface material formed from lava flows generated by large bombardments, on Venus there is evidence of cyclic planet-wide resurfacing and active volcanism, on Earth there are plate tectonics and volcanism, on the Moon the dark basaltic regions were created by lava filling the lowlands, on Mars the northern lowlands are theorized to be fresh lava from a large impact while the southern highlands represent early explosive volcanism, and on Io there is sulfur-rich volcanism induced by tidal forcing from Jupiter.$^{\mathrm{Z16}}$ & It is unknown which resurfacing processes are expected to dominate on large-size (super-Earths) and tidally-locked rocky exoplanets, though if the pattern in the Solar System is to be followed, it is inevitable that some amount of the surface area of any planet will be resurfaced with freshly cooled partial melts from the mantle \citep{Meier2026}.\\
Volcanism  & The process by which interior material is brought to a planetary surface. Provides a mechanism by which mantle-like material ends up on the surface without processes like seafloor spreading or large impacts, and can also provide out-gassing of specific atmospheric species. & Volcanism has been predicted to be the source of atmospheric SO$_2$ on some exoplanets \citep{Bello-Arufe2025}, where the amount of gas melted mantle material can hold is a strong function of its composition \citep{Brugman2021}. It is argued that in planets with high internal heat (i.e., due to tidal forcing) pipe volcanism akin to Io dominates resurfacing \citep{Reinhold2024}, \nt{while models of tidally locked planets predict nightside volcanism to dominate \citep{Meier2026}.}\\
Plate Tectonics  & The process by which crustal material is reprocessed through subduction. In the Solar System this process only occurs on Earth and enables formation of rocks like granites. There are a myriad of other tectonic regimes (where Earth has a mobile lid that causes plate tectonics, and Venus has a proposed episodic lid that causes cyclical planet-wide resurfacing), which are described in \citet{Lyu2025-cf}. The lack of plate tectonics is the stagnant lid regime, which can still have very active geology occurring \citep{Noack2017}.  & As mentioned above in Tertiary Crust, plate tectonics are solely unique to Earth due to surface water, and therefore unlikely to be on hot observable exoplanet surfaces. Other forms of tectonism can still occur though, through processes like episodic tectonics on Venus. It is possible that partially molten lava exoplanets can have regions of solidified crust floating on top of the magma, which can undergo plate tectonic-like processes like subduction \citep{Chao2021}, \nt{or they have single-lid tectonics that recycle solid crust at the terminators \citep{Meier2026}.}  \\
\enddata
\tablecomments{Broad geology definitions and their potential application to exoplanet studies. This table, which is in no way extensive, was generated to help guide future considerations for exogeology studies, foster collaboration between geologists and exoplanet scientists, and contextualize the categorization the surface albedos featured in the \texttt{POSEIDON} surface albedo database. $^{\mathrm{W14}}$ indicates that information preceding the superscript was summarized from \citet{Winter2014}. $^{\mathrm{Z16}}$ indicates that information preceding the superscript was summarized from \citet{Zalasiewicz2016}. $^{\mathrm{C22}}$ indicates a crustal definition from \citet{Condie2022}.}
\end{deluxetable*}

%\clearpage
%\begin{figure*}[ht!]
%    % \vspace{-1.3cm}
%    \includegraphics[width=1.0\textwidth]{esteban-draft-%figure.jpg}
%    \caption{Draft of Esteban Figure }
%\end{figure*}

\clearpage
\setlength{\extrarowheight}{1.2pt}
\startlongtable
\begin{deluxetable*}{lp{4.5cm}p{8.5cm}}
\colnumbers
\tablecolumns{6}
\tablewidth{0pt}
\tabletypesize{\footnotesize}
\tablecaption{Supplementary Geology Categories and Solar System Context \label{table:Surface Albedos Categories}}
\tablehead{Category & Definition & Solar System Context}
\startdata
Minerals & See Table \ref{table:Geology Glossary} & Minerals compose rocks and can be found in their pure form in rocks composed entirely of one mineral (e.g., anorthite and anorthosite, quartz and quartzite), or as large inclusions in rocks. On Earth's surface, minerals formed at high temperatures in the mantle (i.e., olivines, pyroxenes, feldspars) are broken down easily by atmospheric and surface water processes to form the basis of sedimentary rocks while stable minerals, like quartz, aren't.$^{\mathrm{Z16}}$ \\
\arrayrulecolor{lightgray}\hline
$\hookrightarrow$ Ferropericlase & (Mg,Fe)O (also known as magnesiow\"ustite), mineral category that encompasses periclase MgO and w\"ustite FeO & Makes up a portion of the lower mantle of Earth (alongside other lower mantle minerals like perovskite, CaTiO$_3$). While stable at high pressures in the mantle, it is not stable in the low pressures found on the surface.$^{\mathrm{Z16}}$ Due to this, no MgO or FeO albedo data is explicitly included in our database, but we include its definition here due to it being a potentially large mineralogical component of exotic upper mantle compositions \citep{Putirka2019,Putrika2021}.  \\
$\hookrightarrow$ Olivine & (Mg,Fe)$_2$SiO$_4$, mineral category that encompasses the endmembers forsterite Mg$_2$SiO$_4$ to fayalite Fe$_2$SiO$_4$ & Olivine is one of the first minerals in a melt to crystallize out of a magma as a planet initially cools from formation, where due to its high density it often sinks. It can be present in its pure form as large crystals (phenocrysts) in rocks. On Earth's surface it often weathers quickly.$^{\mathrm{Z16}}$ Alongside pyroxene, it represents the most abundant mineral in the upper mantle of Solar System rocky planets \citep{Guimond2024}. Olivine itself is also a large component of the material used to build planetesimals, and therefore has been found ubiquitously in the Solar System \citep{LEONE2023}. \\
$\hookrightarrow$ Pyroxene &  XY(Si,Al)$_2$O$_6$, mineral category that encompasses enstatite MgSiO$_3$, orthopyroxenes (Mg,Fe)$_2$Si$_2$O$_6$, clinopyroxenes Ca(Mg,Fe)Si$_2$O$_6$, and augite & One of the most common minerals that crystallize first out of a magma. Alongside olivine, represents the most abundant mineral in the upper mantle of Solar System rocky planets \citep{Guimond2024}.$^{\mathrm{Z16}}$ %Augite causes a dark grey to black color in basalts. 
\\
$\hookrightarrow$ Silica & SiO$_2$, mineral group composed of multiple crystalline polymorphs (e.g., quartz, tridymite, cristobalite, coesite, opal) & Quartz is the most common polymorph of silica on Earth and comprises about a tenth of Earth's crust, though all polymorphs of silica can be found on Earth in small amounts \citep{Drees1989}. Quartz is stable in the conditions of Earth's surface and therefore is the main component of sand. Glass is amorphous silica.$^{\mathrm{Z16}}$ Other polymorphs of silica can be found on the surfaces of planets; tridymite and cristobalite has been found on Mars and Moon \citep{Seddio2015,Morris2016,Yen2021}, and coesite has been found in impact craters. While not a main mineralogical component of the upper mantle in Solar System planets, it has been found to be a potentially large component of exotic upper mantle compositions \citep{Putrika2021}.\\
\arrayrulecolor{lightgray}\hline
$\hookrightarrow$ Feldspars & Aluminium-silicate mineral category that is composed of alkali feldspars (orthoclase, KaAlSi$_3$O$_8$ to albite, NaAlSi$_3$O$_8$) and plagioclase feldspars (albite. NaAlSi$_3$O$_8$ to anorthite, CaAl$_2$Si$_2$O$_8$) & Feldspar minerals are the most abundant rock-forming mineral in Earth's crust, where the weathering of them forms clays. In primordial Earth and the Moon, plagioclase feldspars condensed out of calcium and aluminum enriched magma (which was depleted of iron and some silica due to olivine condensation) where, due to their low density, the crystallized material floated to the top and became a main component of the primary crust.$^{\mathrm{Z16}}$ \\
\arrayrulecolor{lightgray}\hline
$\hookrightarrow$ Pyrite & FeS$_2$, also known as Fool's Gold & Forms from iron and sulfur reacting together in an oxygen poor environment. It is the most abundant sulfide mineral on Earth, where it primarily forms in oxygen-poor underwater sediments via anaerobic microbes that strip oxygen from sulfate (SO$_\mathrm{4}$$^{\mathrm{-2}}$) ions to produce disulfide (S$^{\mathrm{-2}}_\mathrm{2}$) that reacts with reduced ferrous iron (Fe$^{\mathrm{+2}}$) to form pyrite. Pyrite rapidly decomposes when exposed to the oxidizing environment of Earth's atmosphere.$^{\mathrm{Z16}}$ Pyrite has been found on Mars \citep{Vaniman2014}, hinting that Mars once housed reducing environments.  \\
$\hookrightarrow$ Hematite & Fe$_2$O$_3$ & Forms from iron bonding with oxygen, usually from the weathering of iron-rich materials in oxidizing conditions, or, formed from precipitation in aqueous environments. It ss often found in the soil composition on Earth. When mixed with soils and clays, it can cause a red color; Mars has a thin layer of hematite on its surface causing its famous reddish color.$^{\mathrm{Z16}}$ Hematite has been attributed to being a potential cause of the ferric Fe$^{\mathrm{3+}}$ signal from Venus \citep{Fegley1995}, and has been recently discovered in the highlands of the Moon, despite it lack of a substantial atmosphere \citep{Li2020}.  \\
\arrayrulecolor{black}\hline 
\arrayrulecolor{black}\hline 
Ultramafic Rocks & Category of rocks with chemical compositions that are the most magnesium-rich, silica-poor (SiO$_2$ $\leq$ 45 wt\%) & On Earth, ultramafic rocks were the first to condense out of the primordial magma ocean, forming the upper mantle. These rocks partially melt to form silica-rich magmas that then cool to form mafic rocks.$^\mathrm{Z16}$ \\
$\hookrightarrow$ Peridotite & Rocks made from olivine and pyroxene minerals; by definition, are $>$ 40\% olivine & Peridotites are the main constituent of Earth's upper mantle, where partial melts of peridotite subsequently cool to form basalts. Models have also shown that peridotites are thought to be the main constituent of exoplanet upper mantles \citep{Putrika2021}. Peridotites crystallize from magma early on, and can often found as fragments in mafic rocks as `xenoliths' that form when magma breaks off mantle rocks during its ascent to the surface. On Earth, large dislocated slabs of the lithosphere can be brought to observable crustal layers via tectonics and contain the full strata of basalts, gabbro, peridotite, and spinel (in that order of increasing depth).$^{\mathrm{Z16}}$ The relative abundance of olivine vs pyroxene in peridotite depends on the level of partial melting a rock has experienced, with example rocks in our database being dunite ($>$ 90\% olivine), harzburgite ($<$ 5\% clinopyroxene), and lherzolite (intermediate composition of clino- and ortho-pyroxene).$^{\mathrm{W14}}$\\
$\hookrightarrow$ Pyroxenite & Rocks made from olivines and pyroxenes; by definition, is $>$ 60\% pyroxene & While the olivine and pyroxene minerals in Earth's upper mantle mostly produce peridotite, they produce a small percentage of pyroxenite as well. While it does not dominate any upper mantle in the Solar System, exotic mantle mineralogical compositions could produce abundant pyroxenites in the upper mantle \citep{Putrika2021}. \\
\arrayrulecolor{black}\hline 
\arrayrulecolor{black}\hline 
Mafic Rocks & Broad category of rocks with chemical compositions that are relatively magnesium (Mg) rich (SiO$_2$ $<$ 63 wt\%) & On Earth, mafic rocks form from the cooling of partial melts of ultramafic mantle material, where it forms fresh crustal material at mid-ocean ridges. A majority of Solar System rocky body surfaces are composed of mafic rocks.$^\mathrm{Z16}$\\
$\hookrightarrow$ \textbf{Basalts} (TAS) & Rocks formed from partial melts of ultramafic material, which contain relatively low SiO$_2$ (46-52 wt\%) and low Na$_2$O+K$_2$O ($<$5 wt\%) when compared to other TAS rocks. Basalts have a mineralogical composition comprised mostly of feldspars and pyroxenes, and its intrusive variant is gabbro & Basalts are the most common igenous rock on the surface of Earth and are the main component of Earth's oceanic crust, where fresh crust is created continuously at mid-ocean ridges. They are the most ubiquitous surface component of other Solar System terrestrial planets and moons (Mercury, Venus, southern highlands of Mars, dark lowland regions of the Moon, volanic eruptions of Io) despite different internal structures and large-scale geologic processes in these objects. Most non-exotic silicate mantles are predicted produce basaltic magmas; where, in general, partial melts of ultramafic mantle material will cool and produce a basalt \citep{First2025}.$^\mathrm{Z16}$ The two main sub-categories are alkali basalts, which have higher Na$_2$O+K$_2$O content, and tholeiitic basalts, which are more SiO$_2$ saturated. Tholeiitic basalts are the most common form of basalt on Earth, being the form of basalts that emerge at mid-ocean ridges (they are also the most common form of basalt on the Moon) and represent more shallow mantle melting, while alkali basalts form in regions of low tectonic activity and represent more deeper mantle melting (like volcanic islands and continental rifts).$^{\mathrm{W14}}$ Basalts are usually dark-colored due to augite or pyroxene minerals, but can be lighter due to feldspar minerals. It has been shown that basalt's mid-infrared absorption feature cannot be modeled by mixing individual mineral absorption features, highlighting the necessity of lab data \citep{Hu2012,Ehlmann2012}. \\
\arrayrulecolor{lightgray}\hline
$\hookrightarrow$ Trachybasalt (TAS) & Similar SiO$_2$ content to basalts, but with a higher content of Na$_2$O+K$_2$O (5-7.3 wt\%) & It is found in Earth on some ocean islands, continental rift volcanoes,$^\mathrm{W14}$ and has been found in the Gale Crater on Mars \citep{Edwards2017}. \\
$\hookrightarrow$ Basaltic Andesite (TAS) & Slightly higher SiO$_2$ content to basalts (52-57 wt\%), but with a similar Na$_2$O+K$_2$O content & Found in volcanic arcs (a belt of volcanoes above a subducting plate) on Earth, and have also been found on Mars \citep{Ruff2007,Mullen2013}. \\
$\hookrightarrow$ Tephrite (TAS) & Distinguished from basalt by a lower SiO$_2$ content (41-49 wt\%) and higher Na$_2$O+K$_2$O content (3-9.4 wt\%) & Found in specific regions of Earth and Mars where local magma was enriched in alkaline material.$^\mathrm{W14}$ \\
$\hookrightarrow$ Phonolite (TAS) & Higher SiO$_2$ content than basalts (52.5-61 wt\%), and is the TAS category with the highest amount of Na$_2$O+K$_2$O content ($>$ 12 wt\%) & Rare on Earth, where their formation requires a very low degree of partial melting such that alkali-rich rocks melt but silicon-rich rocks don't.$^\mathrm{W14}$ \\
\arrayrulecolor{lightgray}\hline
$\hookrightarrow$ Gabbro & Is the intrusive variant of basalt, which is more coarse grained & When compared to basalts, which cool rapidly at the surface of fresh oceanic crust due to the magma's interaction with water, gabbro forms in the denser sub-surface portion of oceanic crust on Earth where it can cool slowly (and therefore form larger grains).$^\mathrm{Z16}$ \\
$\hookrightarrow$ Anorthosite & Rock composed of plagioclase feldspar minerals & Anorthsite is a large component of the Moon's surface (bright areas). When plagioclase feldspars condense into anorthosite, they are less dense than the surrounding magma and float to the top, creating a bright, primary crust.$^\mathrm{Z16}$ \\
\arrayrulecolor{black}\hline 
\arrayrulecolor{black}\hline 
Intermediate Rocks \\
$\hookrightarrow$ Andesite (TAS) & Distinguished from basalt by higher SiO$_2$ content (57-73 wt\%) with an identical Na$_2$O+K$_2$O content; its intrusive variant is diorite & Basalts and andesites are often found together, though basalts are more indicative of oceanic crust and ocean floor spreading, while andesite is commonly found in subduction zones (particularly in island arcs) and has been considered a hallmark of Earth's plate tectonics \citep{Tatsumi2015}. Andesitic magma forms from the same mechanism that produces granitic and rhyolitic magma (see Figure \ref{fig:CLR-JWST-sims}), but rapidly cools on the surface (and therefore is not coarsly textured). Andesite has also been found in the crust of Mars, and theorized to be present on Venus \citep{Pavri1992,McSween2009}.\\
$\hookrightarrow$ Trachyte (TAS) &  Higher SiO$_2$ content than basalts (57-69 wt\%) with a large amount of Na$_2$O+K$_2$O content ($>$ 7 wt\%) & Found in specific regions of Earth where alkaline magma has cooled enough (where refractory materials condense and remove themselves from the melt) that the remaining magma has a composition similar to alkali feldspars. Rarely, on Earth, do more alkali-rich magmas form (that generate phonolite).$^\mathrm{W14}$  It has been found in the Gale Crater on Mars \citep{Sautter2016}. \\
\arrayrulecolor{black}\hline 
\arrayrulecolor{black}\hline 
Felsic Rocks & Broad category composing of relatively silicon (Si) rich (SiO$_2$ $\geq$ 63 wt\%) material & On Earth, felsic rocks form from partially melted mafic parent material, typically at subduction zones, where the partial melts are enriched in silica. Earth has the highest percentage of felsic material on the surface when compared to other Solar System planets, due to processes like plate tectonics gradually producing large amounts of continental crust (from repeated processing of oceanic crust) over geologic timescales.$^\mathrm{Z16}$ \\
$\hookrightarrow$ \textbf{Granite} & Rocks formed from the partial melts of mafic material, which contain very high SiO2 ($\geq$ 70 wt\%) and low MgO ($\leq$ 1 wt\%), with a mineralogical composition comprised mostly of quartz and feldspars; extrusive variant is rhyolite & Granite is the main component of Earth's continental crust, where its parent, silica-rich magma is typically formed on Earth by the partial melting of mafic (basaltic) oceanic crust at subduction zones. The silica-rich magma then ascends and cools slowly, forming a coarse grained rock that makes up the continental crust.$^\mathrm{Z16}$ In the Solar System, a widespread presence of granite is unique to Earth due to plate tectonics, though minimal amounts have been found on Mars \citep{Bonin2012}.   \\
\arrayrulecolor{lightgray}\hline
$\hookrightarrow$ Rhyolite (TAS) & Highest amount of SiO$_2$ content ($>$ 69\%) with large range of Na$_2$O+K$_2$O content (0-13 wt\%); intrusive variant is granite & Found wherever silicon-rich lava rapidly cools, and is a common component of volcanic ash (where the silica-rich, viscous nature of rhyolitic magma often results in explosive volcanic eruptions with pyroclastic rocks, on Earth). Forms via the same mechanism as granite, except the parent rhyolitic lava cools rapidly on the surface (forming a fine-grained rock, in lieu of slowly cooled, coarse-grained granite). Obsidian is a rhyolitic glass, pumice is often rhyloitic, and rhyolitic tuffs are a common structure on Earth's surface.$^\mathrm{Z16}$ Rhylotic-like lava has been theorized on Mars due to the discovery of tridymite \citep{Morris2016,Yen2021}. \\
\arrayrulecolor{black}\hline
\arrayrulecolor{black}\hline 
Others \\
$\hookrightarrow$ Clays & Category of sedimentary, soil materials (fine-grained) that forms from clay minerals (hydrous aluminum phyllosilicates) & Clays form in the presence of water. While they are very abundant on Earth$^{\mathrm{Z16}}$, they are relatively rare on other bodies (though clays have been found on Mars, \citealt{Du2023}, and Ceres, \citealt{Rivkin2006}), indicating past or current surface water.  \\
$\hookrightarrow$ Salts & Category of minerals formed from the combination of a positive and negative ion, that commonly precipitate from surface water & The most common salts in Earth's ocean are NaCl (table salt) and MgSO$_4$ (epsom salt) \citep{Log2017}. Solid NaCl has been found in salt deposits on Earth and Mars (indicative of dried-up salt beds) \citep{Bramble2024}, the sub-surface of Europa \citep{Trumbo2019}, and potential salt glacier have been proposed to exist in regions of Mercury \citep{Rodriguez2023}. MgSO$_4$ similarly has been found on Earth and Mars \citep{Vaniman2004}.
\enddata
\tablecomments{Geological categories used to sort the \texttt{POSEIDON} surface albedo database in Table \ref{table:Surface Albedos}. As in Table \ref{table:Geology Glossary}, this table is not extensive and was generated to help guide future considerations for exogelogy studies and foster collaboration between geologists and exoplanet scientists. Basalt and Granite in boldface due to their extensive use in this paper. (1) Category name. The five large categories are Minerals, Mafic Rocks, Intermediate Rocks, Felsic Rocks, and Others. We take our ultramafic, mafic, and felsic definitions from \citet{Paragas2025}. Instead of having a distinct intermediate category, as in \citet{Paragas2025}, we only label rock categories as intermediate if their TAS-defined SiO$_2$ range falls between our mafic and felsic distinguishing SiO$_2$ percentage (63\%). If a sub-category has (TAS) beside it, it means that it is a specific extrusive igenous rock category in the Total Alkali-Silica (TAS) classification diagram where the ranges of SiO$_2$ and alkali metal oxide content for each category are taken from \citet{LeMaitre2005}. We note that that the database presented in this work does not include representive lab data for the following TAS fields: foidite, picrobasalt, basanite, phonotephryte, tephryphonolite, basaltic trachyandesite, trachyandensite, trachydacite, and dacite. (2) Definition of the category. (3) Solar-System Context of the category. $^{\mathrm{W14}}$ indicates that information preceding the superscript was summarized from \citet{Winter2014}. $^{\mathrm{Z16}}$ indicates that information preceding the superscript was summarized from \citet{Zalasiewicz2016}.  }
\end{deluxetable*}

%\begin{deluxetable*}{lp{4cm}p{4cm}p{4cm}}
%\colnumbers
%\tablecolumns{4}
%\tablewidth{0pt}
%\tablecaption{Geology Categories and Exoplanet Detection Inferences \textcolor{red}{WIP}} \label{table:Surface Albedos Categories}
%\tablehead{Category & Formation Inference & Differentiation Inference & Alteration Inference}
%\startdata
%Mafic & If a planet forms magnesium-rich and cools without differentiation, will form primarily mafic rocks on surface. & If a planet differentiates, mafic %rocks can still make up parts of the surface due to volcanism and plate tectonics. & Mafic rocks cannot be made from silicate-rich rocks without melting %(guess).  \\
%$\hookrightarrow$ Basalts & \\
%\hline 
%Felsic\\
%$\hookrightarrow$ Granites & \\
%\hline
%Others \\
%\hline
%$\hookrightarrow$Pyrite & \\
%$\hookrightarrow$Clays & \\
%$\hookrightarrow$Hematite & \\
%\enddata
%\end{deluxetable*}

\startlongtable
\begin{deluxetable*}{lp{4cm}p{9cm}}
\colnumbers
\tablecolumns{3}
\tablewidth{0pt}
\tabletypesize{\footnotesize}
\tablecaption{Detectable Exoplanet Surfaces and Considerations for Interpretations\label{table:Surface Albedos Categories 2}}
\tablehead{Surface Type & Table \ref{table:Surface Albedos Categories} Categories & Potential Considerations for Interpretation if Detected}
\arrayrulecolor{black}
\startdata
Ultramafic Surface & Ferropericlase, Olivine, Pyroxene, Peridotite, Pyroxenite & According to Solar System theory, once a planet has formed its initial surface will be molten and, as it cools, the first material to condense out are dense ultramafic materials that sink forming the upper mantle. In the modern-day Solar System, ultramafic mantle material is partially melted to form mafic (basaltic) material on all rocky planets, though solid ultramafic rocks can be brought to the surface as large inclusions in other rocks (xenoliths) or via dislocated lithospheric slabs (see Peridotite in Table \ref{table:Surface Albedos Categories}). \citet{Hu2012} argue that an ultramafic surface can be an indication of primary crust with mantle overturn, or a secondary crust formed from hot lavas. Mantle overturn is the process wherein largescale convection causes the mantle to essentially `flip' so that dense material rises to top and light mantle material sinks, which could have occurred in the past on Mars, the Earth, the Moon \citep{ElkinsTanton2005,Bedard2018,Liang2024}. The hot lavas mentioned in \citet{Hu2012} are formed by some process like tidal heating, that completely melts ultramafic material (as opposed to partially melting it) in the subsurface and transports the melt to the surface to cool, which is analogous to the ultramafic lavas of Io \citep{Breuer2022}. An ultramafic surface could also be the remnant of a planet's initial formation chemical composition (very high Mg/Si ratio), the result of stripping away of the crust (and revealing the mantle), or dayside reprocessing from intense incident irradiation. \\
\arrayrulecolor{lightgray}\hline
Mafic Surface & \textbf{Basalt}, Trachybasalt, Basaltic Andesite, Tephrite, Phonolite, Gabbro, Anorthosite, Andesite (Intermediate), Trachyte (Intermediate), Basalts in Table \ref{table:Basalt-Library} & From exoplanet mantle modeling and Solar System observations, basalts (and therefore, mafic rocks) are generally thought to be the most common surface component on rocky planets. This is because the secondary crusts of most planets will be basaltic as a natural consequence of what crystallizes (controlled by phase equilibria) from partial melts of ultramafic mantle material (which, from the modeling of rocky exoplanet compositions derived from host star compositions, e.g., \citealt[][]{Putirka2019}, is expected to be the dominate mode of exoplanet materials). Additionally, most Solar System resurfacing processes (see `Resurfacing' in Table \ref{table:Geology Glossary}) \nt{and models for tidally-locked exoplanets \citep[e.g., terminator recycling in][]{Meier2026}} result in mafic surfaces. However, if a planet orbits a Mg-poor star (and therefore can have a more exotic mantle composition, as probed in the \citealt{Putrika2021} white-dwarf pollution study), a mantle could be more mafic (instead of ultramafic), and a mafic surface could indicate processes similar to the ones listed for `Ultramafic Surfaces'. In \citet{Hu2012}, it is argued that a feldspathic surface (i.e., anorthosite) is indicative of a primary crust without mantle overturn (where as the magma ocean cools rocks, like anorthosite, float to the top) while a basaltic surface is indicative of the formation of a a secondary crust. If a mafic rock-type (like basalt) is detected alongside a more felsic rock-type (such as andesite or granite), it can indicate that there are processes such as the remelting of basalts (which occurs due to plate tectonics in subduction zones on Earth) or extensive fractional crystallization (a process which forms felsic material from a mafic parental melt on Earth, Mars, and the Moon,  \citealt{Urdy2018}). Modelers can test for specific Earth-like geologic processes by comparing their data to the basalt library of \citet{First2025}. \\
\arrayrulecolor{lightgray}\hline
Felsic Surface & Silica, Feldspar, Andesite (Intermediate), Trachyte (Intermediate), \textbf{Granite}, Rhyolite & In the Solar System, felsic rocks only compose a large surface component percentage of the Earth, making up the core of continental crust. The buildup of felsic continental crust (over geologic timescales) on Earth has been attributed to the partial melting of mafic, oceanic crust at subduction zones that generate silica-rich magmas \citep[e.g.,][]{Gazel2015}. \citet{Hu2012} argue that granitic surfaces are indicators of a tertiary crust when assuming an Earth-like mantle. However, felsic surfaces are not always an indicator of past or present plate tectonics since exoplanets can form initially silicte-rich, non Solar System-like mantles \citep{Putrika2021}. If a planet's mantle is mafic or felsic, partial melting of the mantle will produce silica-rich magma without plate tectonics. Additionally, metamorphic processes (i.e., rocks subject to high temperatures and high pressures) can form rocks with granite-like compositions (i.e. gneiss) that then can be uncovered on the surface via erosion or active bombardment. \\
\arrayrulecolor{black}\hline
\arrayrulecolor{black}\hline 
Oxidized Surface & Hematite & If a planet has an oxidizing atmosphere, any iron on the surface will be oxidized. Hematite is just one common example of oxidized iron. In \citet{Hu2012}, hematite is indicative of oxidative weathering. Curiously, hematite was recently discovered on the Moon, despite its reducing and thin atmosphere \citep{Li2020}, though this has been attributed to mass transport of oxygen ions from the Earth to the Moon \citep{Zeng2025}. An oxidized surface might also be related to the interior oxygen fugacity of the object (see Oxygen Fugacity in Table \ref{table:Geology Glossary}). \\
\arrayrulecolor{lightgray}\hline
Reduced Surface & Pyrite &  In the solar-system, pyrite is not common as a bulk surface component since it rapidly decomposes when exposed to an oxidizing atmosphere. Therefore, if was detected on an exoplanet it would potentially indicate a largely reducing environment that keeps pyrite stable. In \citet{Hu2012}, it is argued that pyrite (`metal-rich') is indicative of a primary crust with its mantle ripped off, meaning the planet has lost a substantial amount of its crustal and mantle material either due to large impacts (which have been hypothesized to be the cause of the high density population of super-Earths, \citealt{Cambioni2025}), extreme crustal evaporation via incident irradiation (see example, K2-22 b, \citealt{Tusay2025}, or remnant planetary cores around white dwarfs,  \citealt{Veras2019}). A reducing surface might also be related to the interior oxygen fugacity of the object (see Oxygen Fugacity in Table \ref{table:Geology Glossary}). \\
\arrayrulecolor{black}\hline
\arrayrulecolor{black}\hline 
Magma Surfaces & Glasses in Table \ref{table:Lava-Library} & Some tidally locked, rocky exoplanets (like 55 Cancri e) are expected to be hot enough that their daysides host, either full or partial, magma oceans. The composition of these magma oceans can be determined spectrally from their instrinic emission (see \citealt{Fortin2022,Fortin2024}), thin vaporized rock atmospheres \citep{Schaefer2009,Miguel2011,Ito2015,Kite2016,Teske2025}, or thick out-gassed atmospheres \citep{Hu2024} (see Oxygen Fugacity in Table \ref{table:Geology Glossary}). Drawing geologic inferences from the top-level composition of a magma ocean require a framework that includes planetary properties (like gravity), temporal properties (age of planet), atmospheric properties, and water content \citep[see useful review, ][]{Chao2021}. Whether or not a magma ocean does exist on the dayside of a rocky planet, and the magmas ability to dissolve or outgas volatiles, is strong function of incident irradiation recieved, the crustal composition (as different rocks have different melting temperatures and volatile storage properties), and oxygen fugacity \citep{Brugman2021,Gaillard2021,Lin2021b,Meier2026}. If solid crustal material exists surrounding a partial magma ocean, it might be subject to contact metamorphism from neighboring magma, or low-pressure high-temperature metamorphism from incident irradiation (see Metamorphic Rocks in Table \ref{table:Geology Glossary}).  \\
\arrayrulecolor{black}\hline
\arrayrulecolor{black}\hline 
Aqueously Altered Surface & Clays, Salts, some Basalts in Table \ref{table:Basalt-Library} & While many known rocky exoplanets are expected to be too hot on their tidally-locked, observable daysides to harbour surface liquid water, water can alter the surface and leave remnants indicating its past existence. Clays only form on Earth from the interaction of phyllosilicates (like feldspars) and liquid water. Salts are often an indicator of past or present water in the Solar System (i.e., on Earth, Mars, Europa, etc), where they are left as a residue on the surface from evaporating water. Salts are not expected to be stable at high ($>$1700K) temperatures that can exist on the daysides of tidally locked rocky planets, though salt glaciers have been proposed to exist on shadowed regions of Mercury \citep{Rodriguez2023}. Specific minerals, like amphibole and serpentine, can be present in rocks and can indicate that the rock either formed from magma that had large amounts of dissolved water, or the rock interacted with external water, sometime in its past \citep{First2025}.  \\
\hline
\enddata
\tablecomments{Here, we further contextualize the geological categories in Table \ref{table:Surface Albedos Categories} into broad, detectable surface types for rocky exoplanets in the era of JWST, and provide potential pathways to interpret each one if detected. Similar to Tables \ref{table:Geology Glossary} and \ref{table:Surface Albedos Categories}, this table is not extensive and was generated to help guide future considerations for exogeology studies and foster collaboration between geologists and exoplanet scientists. Basalt and Granite in boldface due to their extensive use in this paper. (1) Surface type category, (2) which categories from Table \ref{table:Surface Albedos Categories} fall under each surface type, and (3) potential considerations for how to geologically interpret a surface if detected.}
\end{deluxetable*}

\clearpage
\setlength{\extrarowheight}{1.2pt}
\startlongtable
\begin{deluxetable*}{lllllll}
\colnumbers
\tablecolumns{7}
\tablewidth{0pt}
\tabletypesize{\scriptsize}
\tablecaption{\texttt{POSEIDON} Surface Albedo Database \label{table:Surface Albedos}}
\tablehead{\\ Name & Database & Texture(s) & Sub-Category & SiO$_2$ \% & MgO \% & (Min, Max) \textmu m}
\startdata
\hline
\multicolumn{2}{l}{\textbf{Ultramafic \& Mafic (SiO$_2$ $<$ 63 wt\%)}} \\
\hline
Fe-oxidized$^\mathrm{M}$ & H12 & Powder & Hematite, Basalt & 18.05$^{\mathrm{S}}$ & 2.71$^{\mathrm{S}}$ & (0.30, 25.0) \\
Ice-rich Silicate$^\mathrm{M}$ & H12 & Powder & Basalt & 38.47$^{\mathrm{S}}$ & 5.78$^{\mathrm{S}}$ & (0.30, 25.0) \\
Dunite xenolith & P25 & Crushed, Powder & Olivine-rich peridotite & 39.04 & 49.28 & (0.40, 25.0) \\
Tephrite & H25 & Powder & Tephrite & 41-49$^{\mathrm{T}}$ & N/A  & (0.30, 25.05) \\
Clay$^\mathrm{M}$ & H12 & Powder & Clay & 41.46$^{\mathrm{S}}$ & 30.86$^{\mathrm{S}}$ & (0.30, 25.0) \\
Lunar anorthosite & H25 & Coarse Powder & Anorthosite & 44.08 & 0.09 & (0.28, 49.86)\\
Basalt w/ Olivine Phenocyrsts & P25 & Slab, Crushed, Powder & Basalt & 44.50 & 15.78 & (0.40, 25.0) \\
Lunar mare basalt & H25 & Coarse Powder & Basalt & 44.57 & 11.36 & (0.31, 25.99)\\
Feldspathic & H12 & Powder & Anorthosite & 44.93 & 0.53 & (0.30, 25.0) \\
Harzburgite & H25 & Coarse Powder & Olivine-Pyroxene rich Peridotite & $<$45\%$^{\mathrm{S}}$ & N/A &  (0.30, 25.92)\\
Lherzolite & H25 & Coarse Powder & Pyroxene-rich Peridotite & $<$45\%$^{\mathrm{S}}$ & N/A & (0.30, 25.92)\\
Alkaline Basalt (large) & H25 & Coarse powder & Basalt  & 45-48$^{\mathrm{T}}$ & N/A &  (0.30, 25.05) \\
Alkaline Basalt (small) & H25 & Powder & Basalt & 45-48$^{\mathrm{T}}$ & N/A & (0.30, 25.05) \\
Basalt glass & H25 & Powder & Basalt & 45-48$^{\mathrm{T}}$& N/A & (0.28, 25.05) \\
Basalt tuff  & H25 & Coarse Powder & Basalt (ejecta) & 45-52$^{\mathrm{T}}$ & N/A & (0.30, 25.05)\\
Gabbro & H25 & Crushed & Gabbro & 45-52$^{\mathrm{T}}$ & N/A & (0.30, 49.86) \\
Norite & H25 & Powder & Gabbro & 45-52$^{\mathrm{T}}$ & N/A &  (0.30, 25.05)\\
Mars Basaltic shergottite & H25 & Powder & Basalt (meteoroid) & 45-52$^{\mathrm{T}}$ & N/A & (0.30, 49.86)\\
Mars breccia & H25 & Slab & Basalt (impact regolith) & 45-52$^{\mathrm{T}}$ & N/A & (0.30, 25.05)\\
Trachybasalt & H25 & Powder & Trachybasalt & 45-52$^{\mathrm{T}}$ & N/A & (0.30, 25.05) \\
EG-19-63 Olivine clinopyroxenite & P25 & Slab, Crushed, Powder & Pyroxenite & 47.10 & 24.06 & (0.40, 25.0) \\
EG-19-70 Olivine pyroxenite & P25 & Slab, Crushed, Powder & Pyroxenite & 47.58 & 19.62 & (0.40, 25.0) \\
\textbf{Tholeiitic Basalt} & H25 & Slab & Basalt & 48-52$^{\mathrm{T}}$ & N/A & (0.30, 99.72) \\
Basaltic andesite & P25 & Slab, Crushed, Powder & Basaltic Andesite & 49.24 & 8.53 & (0.40, 25.0) \\
Basaltic & H12 & Powder & Basalt & 50.11 & 4.06 &  (0.30, 25.0) \\
K1919 Basalt & P25 & Slab, Crushed, Powder & Basalt & 50.65 & 6.83 & (0.40, 25.0) \\
Ultramafic$^\mathrm{M}$ & H12 & Powder & Olivine, Pyroxene & 50.72$^{\mathrm{S}}$ & 49.28$^{\mathrm{S}}$ & (0.30, 25.0) \\
EG-16-68 Olivine gabbronorite & P25 & Slab, Crushed, Powder & Olivine, Gabbro & 51.05 & 11.65 & (0.40, 25.0) \\
Phonolite & H25 & Powder & Phonolite & 52.5-61$^{\mathrm{T}}$ & N/A & (0.30, 25.05) \\
STM-101 Andesite & P25 & Slab, Crushed, Powder & Andesite & 52.82 & 4.42 & (0.40, 25.0) \\
\hline
\textbf{True Intermediates} \\
\hline
Diorite & H25 & Powder & Andesite (intrusive var.) & 57-73$^{\mathrm{T}}$ & N/A &  (0.30, 25.05) \\
Andesite & H25 & Slab & Andesite & 57-73$^{\mathrm{T}}$ & N/A & (0.30, 25.92) \\
Trachyte & H25 & Powder & Trachyte & 57.8-69$^{\mathrm{T}}$ & N/A & (0.30, 25.05) \\
\hline
\textbf{Felsic (SiO$_2$ $\geq$ 63 wt\%)} \\
\hline
Dalmatian granite & P25 & Slab, Crushed, Powder & Granite & 67.51 & 0.71 & (0.40, 25.0) \\
Granite & H25 & Powder & Granite & $>$69$^{\mathrm{T}}$ & N/A & (0.30, 25.05) \\
Rhyolite & H25 & Powder & Rhyolite & $>$69$^{\mathrm{T}}$ & N/A & (0.30, 25.05)\\
Orlando gold granite & P25 & Slab, Crushed, Powder & Granite & 73.51 & 0.1 & (0.40, 25.0) \\
\textbf{Granitoid}$^\mathrm{M}$ & H12 & Powder & Granite & 76.66$^{\mathrm{S}}$ & 1.23$^{\mathrm{S}}$ & (0.30, 25.0) \\
\hline
\textbf{Other} \\
\hline
Metal-Rich & H12 & Powder & Pyrite & 0$^\mathrm{S}$ & 0$^\mathrm{S}$ & (0.30, 25.0)   \\
Pyrite & H25 & Powder & Pyrite & 0$^\mathrm{S}$ & 0$^\mathrm{S}$ &  (0.12, 20.02) \\
Hematite (Fe-oxidized) & P25 & Crushed, Powder & Hematite & 2.15 & 0.05 & (0.40, 25.0) \\
Hematite & H25 & Powder & Hematite  & 0$^\mathrm{S}$  & 0$^\mathrm{S}$ &  (0.12, 20.02)\\
Albite (dust) & H25 & Powder & Feldspar & 68.74$^\mathrm{S}$  & 0$^\mathrm{S}$ &  (0.13, 20.02) \\
Magnesium sulfate & H25 & Powder & Salt  & 0$^\mathrm{S}$  & 33.22$^{\mathrm{S}}$ & (0.30, 25.92) \\
\hline
\textbf{Extras} \\
\hline
White & N/A & N/A & A = 1, all \textmu m & N/A & N/A & (0.01, 100.0) \\
Black & N/A & N/A & A = 0, all \textmu m & N/A & N/A & (0.01, 100.0) \\
Red & N/A & N/A & A = 1, 0.620-0.750 \textmu m & N/A & N/A & (0.01, 100.0) \\
Orange & N/A & N/A & A = 1, 0.590-0.620 \textmu m & N/A & N/A & (0.01, 100.0) \\
Yellow & N/A & N/A & A = 1, 0.570-0.590 \textmu m & N/A & N/A & (0.01, 100.0) \\
Green & N/A & N/A & A = 1, 0.495-0.570 \textmu m & N/A & N/A & (0.01, 100.0) \\
Blue & N/A & N/A & A = 1, 0.450-0.495 \textmu m & N/A & N/A & (0.01, 100.0) \\
Purple & N/A & N/A & A = 1, 0.380-0.450 \textmu m & N/A & N/A & (0.01, 100.0) \\
\hline 
\enddata
\tablecomments{The general \texttt{POSEIDON v1.4} database of surface albedos (see Tables \ref{table:Basalt-Library}, \ref{table:Lava-Library}, \ref{table:Surface Albedos Reflections} for more specialized collections). All surface albedos present are in their directional-hemispherical reflectance form and have data in wavelengths necessary to model both surface reflection (UV and visible wavelengths) and emission (infrared wavelengths). (1) Name of the surface albedo sample in the database. $^\mathrm{M}$ indicates that a sample was modeled via mineralogical mixing, otherwise a real sample was measured. We have adopted broad ultramafic + mafic ($\leq$63\% SiO$_2$), and felsic ($>$63\% SiO$_2$) definitions, with the delineating percentage (63\%) from \citet{Paragas2025}, to further sort the table. (2) Original database the data is from; databases are \citet{Hu2012} (H12), \citet{Paragas2025} (P25) and \citet{Hammond2025} (H25). \nt{We note that H12 originally used \citet{Wyatt2001} for basalt, \citet{Cheek2009} for feldspatchic, and the \texttt{\href{www.usgs.gov/labs/spectroscopy-lab/usgs-spectral-library}{USGS
Spectral Library}} \citep{Kokaly2017} for all other data; H25 utilized the \texttt{\href{https://sites.brown.edu/relab/relab-spectral-database/}{RELAB Spectral Database}} \citep{Milliken2021}. Reference numbers for \texttt{RELAB} datasets are in Table 1 in Zenodo Supplementary Material (see \S \ref{sec:zenodo}).} (3) Texture(s) of the sample measured; following the definition of \citet{Paragas2025}, slab is a solid sample, crushed corresponds to coarse crushed grains (500 \textmu m to 1mm grains) and powder corresponds to fine grains (25-80 \textmu m) [note we extend the \citet{Paragas2025} range from 63 to 80 \textmu m to account for many \citet{Hammond2025} samples]. We have also included coarse powders (80-500 \textmu m). (4) Sub-category as given in Table \ref{table:Surface Albedos Categories}. (5, 6) SiO2 and MgO wt\%, where when chemical compositional analysis was performed we report exact values, otherwise we use other methods to report a range: $^\mathrm{S}$ indicates that stoichiometry was utilized to determine wt\%, $^\mathrm{T}$ indicates that the range of compositions was taken from the definition of the rock in the Total Alkali-Silica (TAS) Diagram \citep{LeMaitre2005} (where identical ranges were used for intrusive and glassy variants). (7) Wavelength range of the data. \nt{Plots of albedo vs. wavelength for all entries in the surface albedo database can be seen in `Surface Albedo Database' page in page in \texttt{POSEIDON}'s \href{https://poseidon-retrievals.readthedocs.io/en/latest/content/opacity_database.html}{Opacity Database}.}}
\end{deluxetable*}

\setlength{\extrarowheight}{1.2pt}
\startlongtable
\begin{deluxetable*}{lllll}
\colnumbers
\tablecolumns{5}
\tablewidth{0pt}
\tabletypesize{\footnotesize}
\tablecaption{Natural Basalt Library from \citet{First2025} \label{table:Basalt-Library}}
\tablehead{\\ Name & Texture & Representative Geologic Process & Note & (Min, Max) \textmu m}
\startdata
CNE4 & Slab & Supercontinent break-up & Titanite-rich (5\%) & (1.34, 27.14) \\
CNE6 & Slab & Supercontinent break-up & Pyroxene-rich (5\%) & (1.34, 27.14)  \\
CNE9 & Slab & Supercontinent break-up & Titanite-rich (5\%) & (1.34, 27.14)  \\
GOR2 & Slab & Deep-mantle plume, aqueously altered & Serpentine-rich (25\%) & (1.34, 27.14)  \\
LC02 & Slab & Deep-mantle plume & Fresh, unaltered basalt & (1.34, 27.14)  \\
LC09 & Slab & Deep-mantle plume & Fresh, unaltered basalt & (1.34, 27.14)\\
\textbf{LC10} & Slab & Deep-mantle plume & Fresh, unaltered basalt & (1.34, 27.14)\\
SE04 & Slab & Mid-Ocean Ridge, aqueously altered & Amphibole-rich (50\%) & (1.34, 27.14) \\
\textbf{SE20} & Slab & Mid-Ocean Ridge, aqueously altered & Amphibole-rich (59\%) & (1.34, 27.14)  \\
TABT & Slab & Volcanic-arc basalt & Alumina-rich ($\sim$20\%) & (1.34, 27.14) \\
TAC9 & Slab & Volcanic-arc basalt & Alumina-rich ($\sim$20\%) & (1.34, 27.14) \\
TACA & Slab & Volcanic-arc basalt & Alumina-rich ($\sim$20\%) & (1.34, 27.14) \\
TAPS & Slab & Volcanic-arc basalt & Alumina-rich ($\sim$20\%) & (1.34, 27.14) \\
\textbf{TO01} & Slab & Deep-mantle plume, aqueously altered & Serpentine-rich (61\%) & (1.34, 27.14)  \\
WB04 & Slab & Alkali-rich magma & Basanite & (1.34, 27.14) \\
\hline 
\enddata
\tablecomments{The 15 natural basaltic samples featured in \citet{First2025}, in their directional-hemispherical reflectance form. Basalts are the most abundant volcanic rock on Earth, the Moon, and Mars and are expected to be the most abundant mantle rock in bulk silicate planets around FGMK stars \citep{Putirka2019}. This basaltic library, composed of altered and fresh basalts, can be used to test for specific mineralogical and bulk compositional signals in basalts. In particular, aqueously altered minerals (amphibole and serpentine) present in basalts have strong absorption signals that indicate the presence of past or present surface water, while bulk compositions (e.g., percentage of Al$_2$O$_3$) can probe magmatic processes. Bolded entries were featured in Figure 3 of \citet{First2025}. Prefixes represent where samples were collected: CNE = Coastal New England, GOR = Gorgona, LC = Lunar Crater, SE = Santa Elena, TA = Talamanca, TO = Tortugal, WB = Wards Science. \nt{Plots of albedo vs. wavelength for all entries in the surface albedo database can be seen in `Surface Albedo Database' page in page in \texttt{POSEIDON}'s \href{https://poseidon-retrievals.readthedocs.io/en/latest/content/opacity_database.html}{Opacity Database}.}}
\end{deluxetable*}

\setlength{\extrarowheight}{1.2pt}
\startlongtable
\begin{deluxetable*}{lllll}
\colnumbers
\tablecolumns{5}
\tablewidth{0pt}
\tabletypesize{\footnotesize}
\tablecaption{Lava-World Surface Library from \citet{Fortin2022} \label{table:Lava-Library}}
\tablehead{\\ Name & Texture & Parent Mantle Material & Partial Melt Composition & (Min, Max) \textmu m}
\startdata
Chond-30 & Slab & Peridotite Mantle & 30\% Melt Fraction & (2.5, 28) \\
14684-30 & Slab & Peridotite Mantle & 30\% Melt Fraction & (2.5, 28) \\
65639-30 & Slab & Peridotite Mantle & 30\% Melt Fraction & (2.5, 28) \\
81262-30 & Slab & Peridotite Mantle & 30\% Melt Fraction & (2.5, 28) \\
MM3-10 & Slab & Peridotite Mantle & 10\% Melt Fraction & (2.5, 28)\\
6856-10 & Slab & Peridotite Mantle & 10\% Melt Fraction & (2.5, 28) \\
6856-30 & Slab & Peridotite Mantle & 30\% Melt Fraction & (2.5, 28) \\
6856-50 & Slab & Peridotite Mantle & 50\% Melt Fraction & (2.5, 28) \\
59639-30 & Slab & Pyroxenite Mantle & 30\% Melt Fraction & (2.5, 28) \\
22907-30 & Slab& Pyroxenite Mantle & 30\% Melt Fraction & (2.5, 28) \\
%100963-30 & Slab& Quartz-normative Pyroxenite Mantle & 30\% Melt Fraction & (2.5, 28)  \\
MORB-10 & Slab& Quartz-normative Pyroxenite Mantle & 10\% Melt Fraction & (2.5, 28) \\
MORB-30 & Slab& Quartz-normative Pyroxenite Mantle & 30\% Melt Fraction & (2.5, 28)  \\
MORB-50 & Slab& Quartz-normative Pyroxenite Mantle & 50\% Melt Fraction & (2.5, 28)  \\
MORB-100 & Slab& Quartz-normative Pyroxenite Mantle & 100\% Melt Fraction & (2.5, 28)  \\
%112774-10 & Slab& Quartz-normative Pyroxenite Mantle & 10\% Melt Fraction & (2.5, 28)  \\
\enddata
\tablecomments{The 16 potential lava world surfaces featured in \citet{Fortin2022}, which are directional-hemispherical reflectance measurements of rapidly quenched glasses. The glasses have compositions of partial melts of a parent material (i.e., for Chond-30 the glass that was measured is 100\% composed of material made from the molten component of the parent material that is 30\% melted). Entries are in order of decreasing SiO$_2$ content in the mantle composition they are made from (see Figure 1 of \citealt{Fortin2022}). Includes reference Mid-Ocean Ridge Basalt (MORB), peridotite (Earth-like mantle material, MM3), and chondritic composition (Chond) measurements. Numbered lab samples are partial melts of bulk-silicate planet (BSP, bulk planet minus metallic core) compositions from the exoplanet mantle modeling of \citet{Putirka2019} (where primitive mantle compositions, before crustal formation, were modeled based off of stellar compositions). Peridotite compositions are olivine-rich while pyroxenite compositions are pyroxene-rich. The predicted percentage of exoplanets with each type of mantle depends on metalllic core-formation mechanisms (Mercury-like, Earth-like, and Mars-like): peridotite mantles (26.8\%, 78\%, 95.1\%), pyroxenite mantles (59.8\%, 20.6\%, 4.2\%), and quartz-normative pyroxenite mantles (13.3\%, 1.2\%, 0.2\%). Not represented in this sample are magnesiow\"ustite-normative peridotite mantles, which represents $<$ 0.5\% of predicted exoplanet mantles. See Figure 1 of \citet{Fortin2022} for parent mantle material compositions, and resultant partial melt compositions of the samples. Users should also see \citet{Fortin2024}, which explored in-situ temperature-dependent emissivity measurements. \nt{Plots of albedo vs. wavelength for all entries in the surface albedo database can be seen in `Surface Albedo Database' page in page in \texttt{POSEIDON}'s \href{https://poseidon-retrievals.readthedocs.io/en/latest/content/opacity_database.html}{Opacity Database}.}
%As a fraction of the mantle composition is melted, the bulk composition changes as different materials enter the molten phase, no longer representing the bulk composition of the mantle the magma formed from (Figure 1 in \citet{Fortin2022}).
}
\end{deluxetable*}

\setlength{\extrarowheight}{1.2pt}
\startlongtable
\begin{deluxetable*}{llll}
\colnumbers
\tablecolumns{4}
\tablewidth{0pt}
\tabletypesize{\footnotesize}
\tablecaption{Surface Albedos - Reflection Alone for HWO Applications \label{table:Surface Albedos Reflections}}
\tablehead{Name & Description & Reflectance Type & (Min, Max) \textmu m}
\startdata
\hline
Deciduous & Leaf Piles & Bidirectional reflectance  & (0.302, 14) \\
&& + directional-hemispherical reflectance & \\
Coniferous & Needle Piles & Bidirectional reflectance & (0.302, 14) \\
&& + directional-hemispherical reflectance & \\
White Fig Tree$^{\dagger}$ & Spring Leaves & Bidirectional reflectance & (0.35, 2.5) \\
Basalt & Fresh basalt & Absolute reflectance & (0.30, 2.69)\\
Grass & Grass Mixture & Absolute reflectance &  (0.26, 2.98)\\
Snow & Melting Snow & Absolute reflectance & (0.35, 2.50)\\
Ocean & Ocean Seawater & Absolute reflectance & (0.21, 2.98)\\
Sand & Quartz Sand & Absolute reflectance & (0.24, 2.98) \\
\hline 
Cyanobacteria (MarSipp) & Intertidal microbial mat & Polarized Reflectance & (0.35, 1.10) \\
Cyanobacteria (Drymat) & Dry mat & Polarized Reflectance & (0.35, 1.10) \\
Hypersaline biota (MarGuer) & Hypersaline intertidal microbial mat & Polarized Reflectance & (0.35, 1.10) \\
Purple plankton (Conpurp) & Purple plankton sulfur pool & Polarized Reflectance & (0.35, 1.10) \\
\enddata
\tablecomments{Data originally collected in \citet{GoodisGordin2024} (GG25). Data sourced from the \texttt{\href{https://speclib.jpl.nasa.gov/}{NASA JPL ECOSTRESS Spectral Library}} \citep{Baldridge2009,Meerdink2016,Meerdink2019} and \texttt{\href{www.usgs.gov/labs/spectroscopy-lab/usgs-spectral-library}{USGS
Spectral Library Version 7}} \citep{Kokaly2017}. Microbial mat samples are from \citet{Sparks2021} and are housed at NASA Ames Research Center. Unlike Table \ref{table:Surface Albedos}, albedos included in this database are not standardized to directional-hemispherical reflectances. \texttt{ECOSTRESS} entries are bidirectional in short (0.3-2.5 \textmu m) wavelengths and directional-hemispherical in longer ones. \texttt{USGS} datasets are taken with a myriad of spectrometers that are generally calibrated to absolute reflectances. Reference numbers for \texttt{ECOSTRESS} and \texttt{USGS} datasets are in Table 1 in Zenodo Supplementary Material (see \S \ref{sec:zenodo}). The \citet{Sparks2021} spectra are not explicitly defined, though reflectances are measured with a polarized instrument. $^\dagger$Dataset most similar to `Forest' dataset highlighted in \citet{GoodisGordin2024}. \nt{For more examples of HWO-applicable datasets, see \citet{Zelakiewicz2026}.} \nt{Plots of albedo vs. wavelength for all entries in the surface albedo database can be seen in `Surface Albedo Database' page in page in \texttt{POSEIDON}'s \href{https://poseidon-retrievals.readthedocs.io/en/latest/content/opacity_database.html}{Opacity Database}.}  }
\end{deluxetable*}

% Naming convention for \citet{Coehlo2024}: Name is type of organism / enrichment + color + ID full. Glo, RV, BV, and Ana are bacteria in pure culture (a population of just one species), while E1-E53 are communities made out of different organisms where the dominance is the one naming the sample. "E" means "enrichment" meaning it's a community of several organisms but that community is enriched in either PSB or PNSB.

\begin{acknowledgments}
This research utilizes spectra acquired from the NASA RELAB facility at Brown University, the ECOSTRESS spectral library, the USGS spectral library, and microbial surface spectra collected at NASA Ames Research Center. We would like to express our gratitude to many people and groups for providing their insights and guidance during the development of this manuscript. We thank Kimberly Paragas, Dr. Michael Zhang, Dr. Kenneth Goodis Gordon, Dr. Emily First, and Dr. Marc-Antoine Fortin for help integrating and interpreting surface albedos from their works. We thank Aiden Zelakiewicz for being a beta tester for \texttt{POSEIDON v1.4}, providing useful and constant feedback on the project, \nt{and being the lead author on the `Simulating Habitable Worlds Observatory Reflection Spectra' tutorial notebook}. We thank Dr. Natasha Batalha for providing help in initially developing the code. E.M. acknowledges that this material is based upon work supported by the National Science Foundation Graduate Research Fellowship under Grant No.\ 2139899. \nt{We thank the referees for their very constructive and thorough report of our manuscript that significantly improved the quality of this work.} 
\end{acknowledgments}


\vspace{5mm}
\facilities{JWST}


\software{
POSEIDON \citep{MacDonaldMadhusudhan2017, MacDonald2023,Mullens2024,Mullens2025},
numpy \citep[][]{harris2020array}, SciPy \citep[][]{2020SciPy-NMeth}, matplotlib \citep[][]{Hunter:2007}
%, astropy \citep{2013A&A...558A..33A,2018AJ....156..123A, price2022astropy}.
}



\clearpage
\bibliography{main}{}
\bibliographystyle{aasjournal}



\end{document}